\documentclass[nofootinbib,a4paper,10pt,superscriptaddress,twocolumn,eqsecnum]{revtex4-1}
\usepackage{amsmath,amsfonts,amsthm}
\usepackage{booktabs}
\usepackage[utf8]{inputenc}
\usepackage{siunitx}
\usepackage{times}
\usepackage{pgfplots}
\usepgfplotslibrary{fillbetween}
\pgfplotsset{compat=1.18}
\usetikzlibrary{calc}
\pdfoutput=1
\usepackage{graphicx}
\usepackage{color}
\usepackage{braket}
\usepackage{dcolumn}
\usepackage{bm,url}
\usepackage{subfigure}
\usepackage[usenames,dvipsnames,svgnames]{xcolor}
\usepackage{hyperref}
\hypersetup{colorlinks=true, linkcolor=blue, citecolor=red, urlcolor=Blue, filecolor=magenta}
\definecolor{oxfordblue}{rgb}{0.0, 0.13, 0.28}
\definecolor{burgundy}{rgb}{0.5, 0.0, 0.13}
\definecolor{darkolivegreen}{rgb}{0.33, 0.42, 0.18}
\definecolor{darkblue}{rgb}{0,0,0.5}
\definecolor{richcarmine}{rgb}{0.84, 0.0, 0.25}
\definecolor{bluer}{rgb}{0.00,0.50,0.75}

\begin{document}
	\newcommand{\sigX}{0.03}
	\newcommand{\sigG}{0.10}
	\newcommand\be{\begin{equation}}
		\newcommand\ee{\end{equation}}
	\newcommand\bea{\begin{eqnarray}}
		\newcommand\eea{\end{eqnarray}}
	\newcommand\bseq{\begin{subequations}}
		\newcommand\eseq{\end{subequations}}
	\newcommand\bcas{\begin{cases}}
		\newcommand\ecas{\end{cases}}
	\newcommand{\p}{\partial}
	\newcommand{\f}{\frac}
	
	\title{Testing PPN by the  Rotation of S301 Star}
	\author{Mohsen Khodadi}
	\email{m.khodadi@du.ac.ir}
	\affiliation{School of Physics, Institute for Research in Fundamental Sciences (IPM),	P. O. Box 19395-5531, Tehran, Iran}
	\affiliation{School of Physics, Damghan University, Damghan, 3671645667, Iran}
	\affiliation{Center for Theoretical Physics, Khazar University, 41 Mehseti Str., AZ1096 Baku, Azerbaijan}
	
	\author{Salvatore Capozziello}
	\email{capozziello@na.infn.it}
	\affiliation{Dipartimento di Fisica ``E. Pancini'', Università di Napoli ``Federico II'',	Complesso Universitario di Monte Sant' Angelo, Edificio G, Via Cinthia, I-80126, Napoli, Italy}
	\affiliation{Istituto Nazionale di Fisica Nucleare (INFN), sez. di Napoli, Via Cinthia 9, I-80126 Napoli, Italy}
	\affiliation{Scuola Superiore Meridionale, Via Mezzocannone 4, I-80134, Napoli, Italy.}
	
	\date{\today}
	
	\begin{abstract}
		The recently discovered S301 star, with an orbital period of 8.7 years and
		eccentricity $e = 0.982$, reaches a pericenter of just $140\,R_{\mathrm{s}}$
		--- approximately nine times closer than S2's $\sim 1200R_{\mathrm{s}}$ ---
		offering an unprecedented laboratory for testing
		gravity at large post-Newtonian counting parameters. We
		demonstrate that S301's stellar spin precession provides a novel probe of the
		PPN parameter $\gamma$, scaling as $(2\gamma_{\mathrm{PPN}} + 1)/3$ relative to
		general relativity (GR). For maximum projected velocity shifts of $46.1~{\rm km\,s^{-1}}$, next-generation spectrographs could constrain $\gamma_{\mathrm{PPN}}$ to $\sim10\%$ precision; this forecast is conditional on the treatment of astrophysical systematics. When combined with S301's
		orbital precession --- a substantial $\sim 1.95^\circ$ pericenter advance per
		orbit (GR prediction), approximately an order of magnitude larger than S2's
		$\sim 0.2^\circ$, constraining
		$(2\gamma_{\mathrm{PPN}} - \beta_{\mathrm{PPN}}+2)/3$ --- the spin
		measurement breaks the $\beta$-$\gamma$ degeneracy. A projected MCMC analysis
		of S2 and S301 observations yields
		$\sigma_{2\gamma-\beta}\approx 3\times 10^{-2}$ from the orbital
			precession and $\sigma_{\beta}\approx 3\times 10^{-2}$ (with $\gamma$ fixed to its
			GR value by the spin measurement; $\sigma_\beta\approx0.10$ if the spin uncertainty
			is propagated), with the $\gamma$ precision limited by the spin precession
		measurement to $\sigma_{\gamma}\approx 0.1$.
		 While the spin precession provides an independent consistency
		check, its true significance lies in probing
	the same PPN parameters at unprecedentedly large counting
			parameters, where the non-linear and velocity-dependent PPN terms are
			comparatively enhanced. This complementary approach opens a new window for
		testing the PPN framework
		at large post-Newtonian counting parameters and demonstrates
		the power of combining multiple stellar probes at the Galactic Center.
		
		\vspace{0.5cm}
		
		\textbf{Keywords:} General Relativity, Parametrized Post-Newtonian formalism, Galactic Center, stellar precession
	\end{abstract}
	\maketitle
	
	\section{Introduction}\label{intro}
	
	General relativity (GR), formulated by Einstein in 1915, has withstood over a century of rigorous experimental and observational scrutiny \cite{Will:2014kxa}. At the initial stage of its development, GR was verified and confirmed primarily in the weak gravitational field limit of the Solar System through its classic tests: the perihelion shift of Mercury, the deflection of light by the Sun, confirmed during the 1919 solar eclipse, and the gravitational redshift of spectral lines \cite{mt1973,brumberg1991}. With the advent of modern astronomical observation technologies, however, it has become possible to probe GR predictions in the strong gravitational field regime---a domain where deviations from Einstein's theory might become apparent \cite{Will:2014kxa}.
	Key examples of such strong-gravity probes include:
Binary pulsar timing, where the post-Keplerian parameters extracted from systems like PSR~B1913+16 and the double pulsar PSR~J0737--3039 provide precise tests of strong-field GR, with the latter yielding agreement at the $0.05\%$ level \cite{Hulse:1974eb,Kramer:2006nb,Damour:1991rd,Damour:2014tpa};
Gravitational-wave observations of compact binary mergers, where LIGO/Virgo/KAGRA data enable inspiral-merger-ringdown consistency tests and parameterized deviations from the Kerr metric, all remaining consistent with GR to date \cite{LIGOScientific:2016lio,LIGOScientific:2020tif};
X-ray reflection spectroscopy of accreting black holes, which uses relativistic broadened iron lines and parametric spacetimes to constrain deformation parameters, with current \texttt{relxill\_nk} results favoring the Kerr solution \cite{Bambi:2016sac,Krawczynski:2018fnw,Nampalliwar:2019iti,Abdikamalov:2019hcc};
Black hole shadow imaging by the Event Horizon Telescope, which resolves the near-horizon photon ring of M87* \cite{EventHorizonTelescope:2019ggy} and Sgr~A* \cite{EventHorizonTelescope:2022wkp}, constraining deviations from the GR-based metrics \cite{Volkel:2020xlc,Khodadi:2020gns,Khodadi:2021gbc,Khodadi:2022ulo,Khodadi:2022pqh,Vagnozzi:2022moj,Khodadi:2024ubi,Liu:2025wwq,Capozziello:2014rva,Borka:2015vqa,Dialektopoulos:2018iph};
	and
Strong gravitational lensing by supermassive black holes, where relativistic image positions, Einstein rings, and time delays constrain modified gravity parameters near the photon sphere \cite{Bozza:2010xqn,Keeton:2005jd}.
	Together, these complementary multi-messenger and multi-wavelength probes are steadily charting the strong-field landscape, pushing GR to its observational limits.
	
	The Galactic Center, harboring the supermassive black hole Sagittarius A* (Sgr A*) with a mass of approximately $4.3 \times 10^6$ solar masses \cite{EventHorizonTelescope:2022wkp}, provides a unique and unparalleled laboratory for testing gravity in this strong-field regime. The black hole's proximity to Earth, at a distance of roughly 27,000 light-years, makes it the closest supermassive black hole and the largest in angular size on the sky, with an apparent Schwarzschild radius of 53 microarcseconds \cite{EventHorizonTelescope:2022wkp}. Surrounding Sgr A* is a cluster of high-velocity stars, known as the S-stars, whose trajectories are governed by the gravitational field of the black hole \cite{Vagnozzi:2022moj}. Their orbits, some with periods as short as 16 years and eccentricities approaching unity, have been monitored for decades using adaptive optics and interferometric instruments on 8-10 meter class telescopes, providing a direct probe of spacetime curvature near the event horizon.
	
	In recent years, observations of these S-stars have yielded remarkable confirmations of GR predictions in a regime where the gravitational potential $\varphi/c^2$ and velocities $v/c$ are of order $10^{-4}$, approximately four orders of magnitude stronger than in the Solar System. In 2018, the GRAVITY Collaboration reported the first robust detection of the combined gravitational redshift and relativistic transverse Doppler effect in the orbit of the star S2 during its pericenter passage \cite{GRAVITY:2018ofz}, a result that was independently confirmed by the Keck group \cite{Do:2019txf}.  In 2020, the GRAVITY Collaboration announced the detection of the Schwarzschild precession of S2's orbit---a prograde precession of approximately 12 arcminutes per revolution---consistent with GR predictions at a significance exceeding 7 sigma \cite{GRAVITY:2020gka}. These measurements not only confirmed GR in the strong-field regime but also placed stringent limits on any extended mass distribution within the S2 orbit. The GRAVITY Collaboration established a $1\sigma$ upper limit of approximately $1200\,M_\odot$ for any enclosed mass within S2's orbit, effectively ruling out significant dark matter enhancements in the immediate vicinity of Sgr A* \cite{GRAVITY:2024tth}.
	
	While S2 has been instrumental in establishing the validity of GR at the Galactic Center, its orbital parameters---with a pericenter distance of approximately 1000 Schwarzschild radii---are insufficient to probe the strongest-field effects, such as the black hole's spin or the no-hair theorem \cite{eldayem2026}. The recent discovery of S301, a faint star with a magnitude of $m_K = 19.3$, has opened a new window into this extreme regime \cite{eldayem2026} \footnote{The GRAVITY+ Collaboration recently announced the successful first light and commissioning of its new adaptive optics (AO) system, GPAO (Gravity Plus Adaptive Optics), as detailed in \cite{GRAVITYPlus:2026}. This system constitutes a core element of the broader GRAVITY+ project, which aims to significantly enhance the capabilities of the Very Large Telescope Interferometer (VLTI). During commissioning, GPAO successfully observed a brown dwarf with a K-band magnitude of 21.0. This achievement is particularly relevant for our target, S301, which has a comparable magnitude of $m_K = 19.3$; such faint objects would have been exceedingly difficult to track with previous-generation instrumentation.}. S301 possesses an 8.7-year orbital period and an eccentricity of $e = 0.982$, bringing it within just 140 Schwarzschild radii of the black hole at pericenter---ten times closer than S2 \cite{eldayem2026}. Such an orbit produces a Schwarzschild precession of approximately 2 degrees per revolution, an order of magnitude larger than S2's, and is sensitive to the Lense-Thirring frame-dragging effect, making it a promising target for measuring the spin of Sgr A* \cite{eldayem2026}.
	
	The precise monitoring of stellar dynamics in the Galactic Center enables tests of modified gravity and extended dark mass distributions around Sgr A* \cite{Zakharov:2014kka,Zakharov:2016lzv,DellaMonica:2021xcf,Dialektopoulos:2018iph,Rahman:2018fgy,Zakharov:2018omt,Benisty:2023qcv,Khodadi:2025upl}. A broad program along these lines has been developed in recent	years, using the motion of S2 to constrain modified-gravity models and the dark
	matter distribution near Sgr~A*; see \cite{deLaurentis:2022oqa, Losada:2024wdd,DeMartino:2023qkl,DellaMonica:2023dcw,Fernandez:2023kro,DellaMonica:2022kow,DellaMonica:2022eeg,DellaMonica:2021xcf}.	Dark mass encompasses dark matter, stellar remnants, and other faint objects \cite{Heissel:2021pcw,Lechien:2023psa}. The combined effects of Schwarzschild precession and mass precession on S2's orbit have been used to constrain this dark mass, assuming specific density profiles such as the Navarro-Frenk-White (NFW) or Einasto \cite{Heissel:2021pcw}. Alternatively, model-independent reconstruction using a flexible density profile is possible but requires substantially more observations to achieve comparable precision on mass bounds \cite{Lechien:2023psa}.
	
	Beyond orbital astrometry, the intrinsic rotation of stars on such bound orbits introduces a new and qualitatively different probe of relativistic gravity. In GR, a rotating body in free fall parallel-transports its intrinsic angular momentum along its worldline via Fermi-Walker transport, producing a geodetic precession of its spin axis determined exclusively by the spatial curvature of spacetime \cite{Seoane:2026lyg}. Recently in \cite{Iorio:2025ptl} introduced the concept of a ``nethotron'' to describe any rotating object whose spin axis is cumulatively displaced by the post-Newtonian gravitoelectric and gravitomagnetic components of a massive body's gravitational field \cite{Iorio:2025ptl}. While Gravity Probe B remains the only dedicated test of these effects, Iorio emphasizes that measuring the spin precessions of objects orbiting Sgr A* would provide an additional and largely unexplored resource for constraining the spacetime properties of the central black hole, complementing orbital precession measurements \cite{Iorio:2025ptl}. For an extended rotating star, however, this relativistic signature competes with a classical Newtonian torque generated by the star's oblateness as it moves through the tidal field of the central black hole \cite{Seoane:2026lyg}. What makes S301 extraordinary is the extreme eccentricity of its orbit, which compresses both the relativistic geodetic precession and the Newtonian quadrupole torque into sharply localized step functions at each periapsis passage, while the star's modest radius ensures the classical contamination remains bounded \cite{Seoane:2026lyg}. This enables a clean parametric separation: the geodetic shift scales linearly with the star's rotational velocity, whereas the quadrupole-induced shift depends only on the star's ellipticity and radius \cite{Seoane:2026lyg}.
	
	The Parametrized Post-Newtonian (PPN) formalism provides a unified framework for testing gravitational theories by parameterizing possible deviations from GR through two dimensionless parameters: $\beta$, which measures the nonlinearity in the superposition of gravity, and $\gamma$, which quantifies the amount of space curvature produced per unit mass. Beyond this unifying role, the PPN formalism offers several key practical advantages. It enables direct quantitative comparisons between different gravitational theories and observational data without requiring theory-specific analyses, thereby providing a common language for testing gravity across diverse experiments \cite{Capozziello:2011et,Capozziello:2012ie,Capozziello:2005bu}. It facilitates consistent comparisons across vastly different gravitational regimes---from the weak-field Solar System to the strong-field Galactic Center---allowing results obtained in radically different environments to be meaningfully confronted. Furthermore, it naturally reveals degeneracies between parameters, such as the combination $(2+2\gamma-\beta)/3$ that appears in orbital precession, thereby guiding the design of complementary measurements---like the S301 spin-precession experiment---that can break these degeneracies and yield independent constraints on individual PPN parameters. This versatility has made the PPN framework instrumental in constraining or even excluding specific modified gravity models, from scalar-tensor theories to $f(R)$ gravity and beyond \cite{Sanghai:2016tbi,McManus:2017itv,Bolis:2018kcq,Clifton:2018cef,Alexander:2007vt,Gonzalez-Espinoza:2021nqd,Liodis:2026aka} (see also the review \cite{Joyce:2014kja}). In GR, both PPN parameters equal unity.
	
	While Solar System experiments have constrained these parameters with extraordinary precision---$\gamma$ to $2.3\times10^{-5}$ via Cassini's Shapiro time delay \cite{Bertotti:2003rm} and $\beta$ to $10^{-5}$ via Lunar Laser Ranging \cite{Hofmann:2018myc}---these measurements operate exclusively in the weak-field regime. Current S-star constraints, though consistent with GR, suffer from large uncertainties of order 0.5 due to the inherent degeneracy between $\beta$ and $\gamma$ in orbital precession measurements: the pericenter advance of an S-star only constrains the combination $(2+2\gamma-\beta)/3$. Building on constraints derived from current S-star orbits \cite{Gainutdinov:2020bbv}, future observations combining S2 and S62 are forecast to significantly improve the precision of $\beta_{\rm PPN}$ and $\gamma_{\rm PPN}$ measurements, achieving $\sim 1$--$2\%$ precision on both parameters \cite{Losada:2024wdd}.
	
	The S301 spin-precession experiment proposed in this work offers a complementary approach: by measuring the geodetic precession of the star's rotational axis, it directly constrains $\gamma$ alone. Combined with S301's orbital precession, this yields individual constraints on both PPN parameters in a previously inaccessible strong-field regime \cite{Seoane:2026lyg}. Beyond constraining the PPN parameters, S301 also offers a unique opportunity to measure the spin of Sgr A* through the Lense-Thirring effect. A companion paper \cite{Piran:2026zjs} demonstrates that while S301's Lense-Thirring nodal precession exceeds that of S2 by a factor of approximately 30, the Newtonian nodal precession induced by any extended mass distribution remains comparable for stars with similar apocenters. By using S2, S55, and S38---stars that share S301's apocenter but have much larger pericenters---as calibrators of the Newtonian background, the spin signal can be isolated. This joint strategy, combined with the time-variation discriminant from S301's rapid Schwarzschild apsidal advance, brings the measurement of both the magnitude and direction of Sgr A*'s spin within reach of continued GRAVITY+ astrometry and ELT spectroscopy. These complementary approaches demonstrate that S301 is a uniquely powerful tool for testing different aspects of GR in the strong-field regime---the PPN parameters via spin precession, and the black hole's spin via Lense-Thirring precession. This endeavor is further motivated by recent work showing that disk-like extended mass distributions can induce precessions competitive with the Lense-Thirring effect for S301, underscoring the need to constrain such structures for an unbiased spin measurement \cite{Foschi:2026osv}.
	
	
	The remainder of this paper is structured as follows. Section \ref{ppn} introduces the PPN formalism, deriving the metric, Lagrangian, and equations of motion from the Schwarzschild solution. Section \ref{s301} presents the spin-precession observable, deriving the geodetic scaling relation for \(\gamma_{\mathrm{PPN}}\) and constraints from projected velocity shifts. Section~\ref{beta} addresses the complementary orbital dynamics, showing how S301's substantial pericenter advance ($\sim 1.95^\circ$ per orbit) constrains the $\beta$-$\gamma$ combination and breaks the degeneracy when combined with spin precession. Section \ref{Re} compares our results with existing and future constraints across gravitational regimes. Section \ref{sec:analysis} presents a MCMC analysis quantifying the projected precision from combining S2, and S301. Finally, Section \ref{con} discusses broader implications, observational requirements, and concluding remarks.
	
	\section{The PPN Framework}\label{ppn}
	We establish the PPN framework for analyzing S-star orbits around Sgr A* by starting from the standard Schwarzschild solution of Einstein's field equations:
	
	\begin{equation}\label{tag1}
		ds^2 = \left(1-\frac{2GM}{c^2r}\right)c^2 dt^2 - \frac{dr^2}{\left(1-\frac{2GM}{c^2r}\right)} - r^2 d\Omega^2,
	\end{equation}
	
	This metric describes the spacetime geometry around a spherically symmetric, non-rotating mass in standard Schwarzschild coordinates $(t,r,\theta,\phi)$, where the radial coordinate $r$ has the geometric meaning that a circle of constant $r$ has circumference $2\pi r$, though $r$ is not the proper distance to the center \cite{mt1973}. For practical relativistic celestial mechanics, it is transformed to isotropic coordinates $(t,\rho,\theta,\phi)$ using \cite{Gainutdinov:2020bbv, brumberg1991}:
	\begin{equation}\label{tag2}
		r = \rho\left(1 + \frac{GM}{2c^2\rho}\right)^2
	\end{equation}
	This transformation makes the spatial part of the metric conformally Euclidean, allowing Cartesian coordinates $\mathbf{x}$ with $\rho = |\mathbf{x}|$
	\begin{equation}\label{tag3}
		ds^2 = \frac{\left(1-\frac{GM}{2c^2|\mathbf{x}|}\right)^2}{\left(1+\frac{GM}{2c^2|\mathbf{x}|}\right)^2}c^2 dt^2 - \left(1+\frac{GM}{2c^2|\mathbf{x}|}\right)^4 d\mathbf{x}^2
	\end{equation}
	Introducing the Newtonian potential $\varphi_N = -GM/\rho = -GM/|\mathbf{x}|$
	\begin{equation}\label{tag4}
		ds^2 = \frac{\left(1+\frac{\varphi_N}{2c^2}\right)^2}{\left(1-\frac{\varphi_N}{2c^2}\right)^2}c^2 dt^2 - \left(1-\frac{\varphi_N}{2c^2}\right)^4 d\mathbf{x}^2
	\end{equation}
	Expanding to first Post-Newtonian order ($g^{00}$ to $O(c^{-6})$ and $g^{xx}$ to $O(c^{-4})$) yields the PN Schwarzschild metric
	\begin{equation}\label{tag5}
		ds^2 = \left(1 + \frac{2\varphi_N}{c^2} + \frac{2\varphi_N^2}{c^4}\right)c^2 dt^2 - \left(1 - \frac{2\varphi_N}{c^2}\right)d\mathbf{x}^2 + O(c^{-6})
	\end{equation}
	Generalizing to the PPN formalism introduces the parameters $\beta_{\rm PPN}$ and $\gamma_{\rm PPN}$, which measure nonlinearity in gravity superposition and space curvature per unit mass, respectively \cite{Will:2014kxa}:
	\begin{align}\label{tag6}
		ds^2 = \left(1 + \frac{2\varphi_N}{c^2} + \beta_{\rm PPN}\frac{2\varphi_N^2}{c^4}\right)c^2 dt^2 - \nonumber \\
		\left(1 - \gamma_{\rm PPN}\frac{2\varphi_N}{c^2}\right)d\mathbf{x}^2 + O(c^{-6})
	\end{align}
	Dividing by $c^2dt^2$ and expanding:
	\begin{equation}\label{tag7}
		\frac{1}{c^2}\left(\frac{ds}{dt}\right)^2 = 1 - \frac{\dot{\mathbf{x}}^2}{c^2} + \frac{2\varphi_N}{c^2} + \gamma_{\rm PPN}\frac{2\varphi_N\dot{\mathbf{x}}^2}{c^4} + \beta_{\rm PPN}\frac{2\varphi_N^2}{c^4} + O(c^{-6})
	\end{equation}
	Taking the square root with accuracy $O(c^{-6})$ gives:
	\begin{align}\label{tag8}
		\frac{1}{c}\frac{ds}{dt}& = 1 - \frac{\dot{\mathbf{x}}^2}{2c^2} + \frac{\varphi_N}{c^2} - \frac{\dot{\mathbf{x}}^2}{8c^4} + (1+2\gamma_{\rm PPN})\frac{\varphi_N\dot{\mathbf{x}}^2}{2c^4} + \nonumber \\
		&(2\beta_{\rm PPN}-1)\frac{\varphi_N^2}{2c^4} + O(c^{-6})
	\end{align}
	From the variational principle $\delta\int ds = \delta\int (ds/dt)dt = 0$, multiplying by $-c^2$ and dropping constant and $O(c^{-6})$ terms yields the PPN Lagrangian:
	\begin{align}\label{tag9}
		L &= \frac{\dot{\mathbf{x}}^2}{2}\left(1 + \frac{\dot{\mathbf{x}}^2}{4c^2} - (1+2\gamma_{\rm PPN})\frac{\varphi_N}{c^2}\right) - \nonumber \\
		&	\varphi_N\left(1 + (2\beta_{\rm PPN}-1)\frac{\varphi_N}{2c^2}\right)
	\end{align}
	The Euler-Lagrange equations then produce the PPN equations of motion:
	\begin{align}\label{tag10}
		\ddot{\mathbf{x}} =& -\nabla\varphi_N\left(1 + 2(\beta_{\rm PPN}+\gamma_{\rm PPN})\frac{\varphi_N}{c^2} + \gamma_{\rm PPN}\frac{\dot{\mathbf{x}}^2}{c^2}\right) + \nonumber \\
		&	(2\gamma_{\rm PPN}+2)\left(\nabla\varphi_N\cdot\frac{\dot{\mathbf{x}}}{c}\right)\frac{\dot{\mathbf{x}}}{c}
	\end{align}
	Finally, the GR limit corresponds to $\beta_{\rm PPN} =1= \gamma_{\rm PPN} $:
	\begin{equation}\label{tag11}
		\ddot{\mathbf{x}} = -\nabla\varphi_N\left(1 + 4\frac{\varphi_N}{c^2} + \frac{\dot{\mathbf{x}}^2}{c^2}\right) + 4\left(\nabla\varphi_N\cdot\frac{\dot{\mathbf{x}}}{c}\right)\frac{\dot{\mathbf{x}}}{c}
	\end{equation}
	This framework provides the foundation for the S301 analysis.

Before proceeding, a further explanation is necessary here. The PPN formalism is a weak-field, slow-motion expansion in the counting parameters $\varphi_N/c^2$ and $v^2/c^2$, truncated at first post-Newtonian order.
At pericenter, S301 reaches $r_p\approx140\,R_s$, with
		\begin{equation}
			\frac{\varphi_N}{c^2}\sim\frac{GM}{c^2 r_p}\sim10^{-4},
			\qquad
			\frac{v_p}{c}\sim0.08,
			\qquad
			\frac{v_p^2}{c^2}\sim6\times10^{-3}.
		\end{equation}
		These are roughly four orders of magnitude larger than the corresponding Solar
		System values ($\varphi_N/c^2\sim10^{-8}$, $v^2/c^2\sim10^{-8}$) and about an order
		of magnitude larger than those of S2 at pericenter
		($\varphi_N/c^2\sim10^{-5}$, $v/c\sim0.03$). At these values the post-Newtonian
		expansion remains convergent, and the leading corrections retained in
		Eqs.~(\ref{tag5})--(\ref{tag10}) are under control; the next-to-leading terms are suppressed
		by additional powers of $v^2/c^2\sim10^{-2}$ and are negligible at the
		$\sim10\%$ precision forecast for $\gamma_{\rm PPN}$ and at the
		$\sim3\times10^{-2}$ precision forecast for $2\gamma_{\rm PPN}-\beta_{\rm PPN}$.
		The S301 measurement therefore probes the same standard PPN parameters
		$\beta_{\rm PPN}$ and $\gamma_{\rm PPN}$ that are constrained in the Solar
		System, but at PPN counting parameters four orders of magnitude larger.

	\section{The S301 Spin-Precession Observable}\label{s301}
	
	Let us propose geodetic precession in the PPN framework. For a gyroscope in the PPN metric given by Eq.~(\ref{tag6}), the geodetic precession vector is \cite{Will:2014kxa, mt1973}:
	\begin{equation}
		\boldsymbol{\Omega}_{\text{geod}}^{\text{(PPN)}} = \left(\frac{2\gamma_{\rm PPN} + 1}{2}\right)\frac{GM}{c^2 r^2}(\mathbf{n} \times \mathbf{v})
		\label{eq:ppn_geodetic}
	\end{equation}
	where $\mathbf{n} = \mathbf{r}/r$ is the unit radial vector and $\mathbf{v} = \dot{\mathbf{x}}$. In GR ($\gamma_{\rm PPN} = 1$), this gives the familiar $3/2$ coefficient, i.e.,
	\begin{equation}
		\boldsymbol{\Omega}_{\text{geod}}^{\text{(GR)}} = \frac{3GM}{2c^2 r^2}(\mathbf{n} \times \mathbf{v})
		\label{eq:gr_geodetic}
	\end{equation}
The S301 star orbiting Sgr A* provides a unique laboratory for testing GR at PPN counting parameters four orders of magnitude larger than those accessible in the Solar System, as already discussed. Its extreme eccentricity ($e = 0.982$) localizes the precessional evolution to discrete step functions at periapsis, enabling a clean separation of relativistic and classical effects \cite{Seoane:2026lyg}. The key observable is the absolute projected rotational velocity shift:
	
	\begin{equation}
		|\Delta v \sin i| \equiv |\Delta(v_{\text{rot}} \sqrt{1-(\mathbf{s}\cdot\hat{\mathbf{o}})^2})|
		\label{eq:observable}
	\end{equation}
	
	where $\mathbf{s}$ is the unit spin vector and $\hat{\mathbf{o}}$ is the line-of-sight unit vector.
	
	For a small angular displacement $\Delta i$ of the spin axis:
	
	\begin{equation}
		\Delta(v \sin i) \approx v_{\text{rot}} \cos i \, \Delta i
		\label{eq:small_angle}
	\end{equation}
	
	At periapsis, $\mathbf{v} \perp \mathbf{n}$, so $|\mathbf{n} \times \mathbf{v}| = v_p$. Substituting $r_p = a(1-e)$ and $v_p = \sqrt{GM(1+e)/(a(1-e))}$ into Eq.~(\ref{eq:ppn_geodetic}):
	\begin{align}
		|\boldsymbol{\Omega}_{\text{geod},p}^{\text{(PPN)}}| &= \frac{(2\gamma_{\rm PPN}+1)}{2}\frac{GM}{c^2 r_p^2} v_p \nonumber \\
		&= \frac{(2\gamma_{\rm PPN}+1)}{2}\frac{GM}{c^2 [a(1-e)]^2} \sqrt{\frac{GM(1+e)}{a(1-e)}} \nonumber \\
		&= \frac{(2\gamma_{\rm PPN}+1)}{2}\frac{(GM)^{3/2}}{c^2} a^{-5/2}(1+e)^{1/2}(1-e)^{-5/2} \label{eq:ppn_peak}
	\end{align}
	As a result, the ratio to GR reads off
	\begin{equation}
		\frac{|\boldsymbol{\Omega}_{\text{geod},p}^{\text{(PPN)}}|}{|\boldsymbol{\Omega}_{\text{geod},p}^{\text{(GR)}}|} = \frac{2\gamma_{\rm PPN}+1}{3}
		\label{eq:ppn_ratio}
	\end{equation}
	This is the key scaling relation. Note that $\beta_{\rm PPN}$ does not appear in the geodetic precession at this order; it only affects the orbital dynamics through the equations of motion.
	
	\begin{table}[h]
		\centering
		\caption{Orbital parameters of S301 \cite{eldayem2026, Seoane:2026lyg}.}
		\label{tab:params}
		\begin{tabular}{lcc}
			\hline
			Parameter & Symbol & Value \\
			\hline
			Orbital period & $P$ & 8.7 years \\
			Eccentricity & $e$ & 0.982 \\
			Black hole mass & $M$ & $4.3 \times 10^6 M_\odot$ \\
			Gravitational radius & $GM/c^2$ & $\approx 6.4 \times 10^9$ m \\
			Schwarzschild radius & $R_s = 2GM/c^2$ & $\approx 0.08$ AU \\
			Semi-major axis & $a$ & $\approx 7000\,R_s$\footnote{The semi-major axis of $a \approx 7000 R_{\rm s}$ follows from the orbital period via Kepler's third law: $a = \left( \frac{G M P^2}{4\pi^2} \right)^{1/3}.$} \\
			Periapsis distance & $r_p = a(1-e)$ & $\approx 140\,R_s $ \\
			Periapsis velocity & $v_p$ & $\approx 25,000$ km/s \\
			\hline
		\end{tabular}
	\end{table}
	
	\subsection{Velocity shift scaling and observational constraints}
	
	The geodetic angular displacement per orbit scales as:
	\begin{equation}
		\Delta i_{\text{geod}}^{\text{(PPN)}} = \frac{2\gamma_{\rm PPN}+1}{3} \Delta i_{\text{geod}}^{\text{(GR)}}
		\label{eq:ppn_delta_i}
	\end{equation}
	The GR prediction corresponds to $\gamma_{\rm PPN} = 1$, giving $\Delta v_{\text{geod}}^{\text{(GR)}}$. From Eq.~(\ref{eq:small_angle}), $\Delta v \approx v_{\text{rot}} \cos i \, \Delta i$. Substituting Eq.~(\ref{eq:ppn_delta_i}) gives
	\begin{equation}
		\Delta v_{\text{geod}}^{\text{(PPN)}} \approx \frac{2\gamma_{\rm PPN}+1}{3} v_{\text{rot}} \cos i \, \Delta i_{\text{geod}}^{\text{(GR)}}
		\label{eq:ppn_delta_v}
	\end{equation}
	The total observable including the classical quadrupole term (from \cite{Seoane:2026lyg}) is
	\begin{equation}
		|\Delta v\sin i|^{\mathrm{(PPN)}} = \left|\frac{2\gamma_{\mathrm{PPN}} + 1}{3}\Delta v_{\mathrm{geod}}^{\mathrm{(GR)}} + \Delta v_{\mathrm{quad}}\right|
		\label{eq:total_observable}
	\end{equation}
	where $\Delta v_{\mathrm{quad}}$ is the classical quadrupole-induced velocity shift computed in Ref. \cite{Seoane:2026lyg}. In the forecast below we treat $\Delta v_{\mathrm{quad}}$ as a systematic with an estimated uncertainty, rather than as a known deterministic contribution; this is discussed further in Sec.~\ref{sec:systematics}.
	
	From the Monte Carlo simulations in \cite{Seoane:2026lyg} we provide Table \ref{tab:shifts}.
	\begin{table}[h]
		\centering
		\caption{Expected shifts from S301. The GR prediction is the central value of the spin-precession velocity shift; the resolution is the projected spectroscopic uncertainty of the specified instrument generation. The two are independent quantities and are listed in separate columns.}
		\label{tab:shifts}
		\begin{tabular}{lcc}
			\hline
			Configuration & GR Prediction & Resolution \\
			\hline
			Median shift & $3-6.3$ km/s & Next-gen: $3$ km/s \\
			Maximum shift & $46.1$ km/s & Current: $50$ km/s \\
			\hline
		\end{tabular}
	\end{table}
Given the PPN-modified velocity shift from Eq.~(\ref{eq:ppn_delta_i}), if $\gamma_{\rm PPN} \neq 1$, the PPN prediction differs from GR. The deviation is:
	\begin{align}
		\Delta v_{\text{geod}}^{\text{(PPN)}} - \Delta v_{\text{geod}}^{\text{(GR)}}
		&= \left(\frac{2\gamma_{\rm PPN}+1}{3} - 1\right) \Delta v_{\text{geod}}^{\text{(GR)}} \nonumber \\
		&= \frac{2(\gamma_{\rm PPN} - 1)}{3} \Delta v_{\text{geod}}^{\text{(GR)}}
		\label{eq:deviation}
	\end{align}
	If the experiment measures the velocity shift and finds it consistent with GR within the instrumental resolution $\sigma_v$, then:
	\begin{equation}
		|\Delta v_{\text{geod}}^{\text{(PPN)}} - \Delta v_{\text{geod}}^{\text{(GR)}}| < \sigma_v
		\label{eq:observational_constraint}
	\end{equation}
	Substituting Eq.~(\ref{eq:deviation}) into Eq.~(\ref{eq:observational_constraint}):
	
	\begin{equation}
		\left|\frac{2(\gamma_{\rm PPN} - 1)}{3}\right| \Delta v_{\text{geod}}^{\text{(GR)}} < \sigma_v
	\end{equation}
	
	Thus, the general constraint on $\gamma_{\rm PPN}$ is:
	
	\begin{equation}
		|\gamma_{\rm PPN} - 1| < \frac{3\sigma_v}{2\,\Delta v_{\text{geod}}^{\text{(GR)}}}
		\label{eq:gamma_constraint}
	\end{equation}
	From the Monte Carlo simulations in \cite{Seoane:2026lyg}, the median GR shift is $\Delta v_{\text{geod}}^{\text{(GR)}} \approx 5$ km/s. With next-generation spectrographs achieving $\sigma_v \approx 3$ km/s:
	\begin{align}
		|\gamma_{\rm PPN} - 1|<  0.9 ~~\Longrightarrow~~ 0.1 < \gamma_{\rm PPN} < 1.9
		\label{eq:gamma_med}
	\end{align}
	For the maximum shift case, $\Delta v_{\text{geod}}^{\text{(GR)}} = 46.1$ km/s:
	\begin{align}
		|\gamma_{\rm PPN} - 1| &< \frac{9}{92.2} \approx 0.0976 \\
		-0.0976 &< \gamma_{\rm PPN} - 1 < 0.0976
	\end{align}
	which results in
	\begin{equation}
		0.9 < \gamma_{\rm PPN} < 1.1
		\label{eq:gamma_max_compact}
	\end{equation}
	
	\begin{table}[h]
		\centering
		\caption{Comparison of PPN parameter constraints across different regimes. All entries are $1\sigma$ constraints unless otherwise stated; the S301 entry is the conditional forecast of Sec.~\ref{sec:systematics}.}
		\begin{tabular}{lccc}
			\hline
			Experiment & Regime & $\beta_{\rm PPN}$ & $\gamma_{\rm PPN}$  \\
			\hline
			Cassini \cite{Bertotti:2003rm} & Solar System & --- & $10^{-5}$  \\
			Lunar Laser Ranging \cite{Hofmann:2018myc} & Solar System & $10^{-5}$ & $10^{-4}$ \\
			S2, S38, S55 \cite{Gainutdinov:2020bbv} & Galactic Center & $0.97^{+0.42}_{-0.65}$ & $0.81^{+0.46}_{-0.60}$  \\
			S2 + S62 (future) \cite{Losada:2024wdd} & Galactic Center & $\sim 2\%$ & $\sim 1\%$  \\
			S301 (projected) This work & Galactic Center & $\sim 10\%$ & $\sim 10\%$   \\
			\hline
		\end{tabular}
		\label{tab:comparison}
	\end{table}
	
	\section{Constraints on $\beta_{\rm PPN}$ from Orbital Dynamics}\label{beta}
	
	The full PPN equations of motion (Eq.~\ref{tag10}) produce a pericenter precession rate that depends on both $\beta_{\rm PPN}$ and $\gamma_{\rm PPN}$. The standard formula for pericenter precession per orbit is \cite{Will:2014kxa}:
	\begin{equation}
		\Delta \omega = \frac{6\pi GM}{c^2a(1 - e^2)}\times \frac{2 + 2\gamma_{\mathrm{PPN}} - \beta_{\mathrm{PPN}}}{3}.
		\label{eq:ppn_precession}
	\end{equation}
	In GR ($\beta_{\rm PPN} = \gamma_{\rm PPN} = 1$), this gives
	\begin{equation}
		\Delta \omega_{\mathrm{GR}} = \frac{6\pi GM}{c^2a(1 - e^2)}
		\label{eq:gr_precession}
	\end{equation}
	By taking values listed in Table \ref{tab:params} ($r_p \approx 140R_{\mathrm{s}}$ and $e=0.982$), the semi-major axis in units of $GM/c^2$ is: $a = \frac{r_p}{1-e} \approx 15556 \, GM/c^2$. The orbital advance of S301 due to GR is then:
	\begin{equation}
		\Delta \omega_{\mathrm{GR}} = \frac{6\pi}{15556 \times (1 - 0.982^2)}
		\approx 0.0340\ \mathrm{rad}
		\approx 1.95^{\circ}
		\label{eq:gr_precession_numeric}
	\end{equation}
	This substantial precession per orbit is approximately an order of magnitude larger than S2's precession ($\sim 0.2^{\circ}$ per orbit) \cite{GRAVITY:2020gka}.
	
	The observed pericenter precession would be:
	\begin{equation}
		\Delta \omega_{\mathrm{obs}} = \frac{2 + 2\gamma_{\mathrm{PPN}} - \beta_{\mathrm{PPN}}}{3}\Delta \omega_{\mathrm{GR}}
		\label{eq:obs_precession}
	\end{equation}
	If S301's orbital precession is measured and matches GR, then:
	\begin{align}
		\frac{2 + 2\gamma_{\mathrm{PPN}} - \beta_{\mathrm{PPN}}}{3} = 1
		\;\Longrightarrow\;
		\beta_{\mathrm{PPN}} - 2\gamma_{\mathrm{PPN}} = -1
		\label{eq:beta_gamma_relation}
	\end{align}
	recovers the GR relation. If a deviation from GR is detected, the combination $(2+2\gamma-\beta)/3$ would deviate from unity.
	
	If both the spin-precession and orbital-precession measurements are performed and found consistent with GR, then the combined constraints would be:
	\begin{align}\label{den}
		&\gamma_{\mathrm{PPN}}\in [0.9,1.1]\quad (\text{from spin precession}) \\
		&\beta_{\mathrm{PPN}} - 2\gamma_{\mathrm{PPN}} = -1\quad (\text{from orbital precession, assuming GR})
		\label{denn}
	\end{align}
	which, for $\gamma_{\mathrm{PPN}}\in[0.9,1.1]$, yields
	\begin{equation} \label{bet}
		\beta_{\mathrm{PPN}} = 2\gamma_{\mathrm{PPN}} - 1 \in [0.8,1.2].
	\end{equation}
	The key point is that $\beta_{\rm PPN}$ is not independently constrained by the spin-precession measurement; it only gets constrained when combined with the orbital precession measurement under the assumption that GR holds for the orbital dynamics.
	
	\begin{figure}[ht!]
		\includegraphics[width=0.5\textwidth]{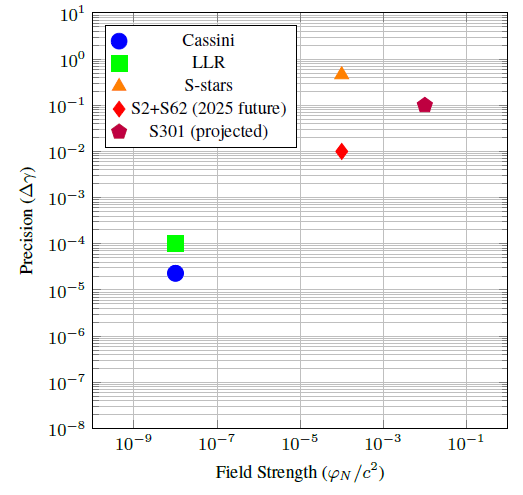}
		\caption{Comparison of constraints on the standard PPN parameter
			$\gamma_{\rm PPN}$ across different values of the PPN counting parameter	$\phi_N/c^2$. Solar System tests provide the highest precision but the smallest		counting parameters. S301 provides lower precision $(\sim10\%)$ but probes
			$\phi_N/c^2\sim10^{-4}$ and $v/c\sim0.08$, four orders of magnitude larger than		in the Solar System. All points refer to the same PPN parameter $\gamma_{\rm PPN}$.}
		\label{fig:comparison}
	\end{figure}
	
	\section{Comparison with other Results}\label{Re}
	
	To contextualize the projected constraints from the S301 spin-precession experiment, it is instructive to compare them with existing and future measurements of the PPN parameters $\beta$ and $\gamma$ across different gravitational regimes. The PPN formalism provides a unified framework for testing gravitational theories by parameterizing possible deviations from general relativity. The parameters $\beta$ and $\gamma$ have been constrained with extraordinary precision in the Solar System through experiments such as the Cassini spacecraft's measurement of the Shapiro time delay and lunar laser ranging (LLR). These experiments achieve relative precisions of $10^{-4}$ to $10^{-5}$ for $\beta$ and $\gamma$, but they operate in the weak-field regime where $\varphi/c^2 \sim 10^{-8}$ and $v^2/c^2 \sim 10^{-8}$. In contrast, the Galactic Center provides a fundamentally different testing ground, with S-stars reaching $\varphi/c^2 \sim 10^{-4}$ and $v/c \sim 0.01$, approximately four orders of magnitude stronger than the Solar System. Current constraints from S-star orbits, while consistent with general relativity, suffer from large uncertainties of order $0.5$ due to the limited observational baseline and the inherent degeneracy between $\beta$ and $\gamma$ in orbital precession measurements. Future observations combining S2's orbital precession with S62's gravitational lensing are projected to achieve $\sim 1-2\%$ precision on both parameters by breaking this degeneracy.
	
	The S301 experiment proposed in this work would provide a complementary measurement in an even stronger field regime, with $r_p\sim 140R_{\mathrm{s}}$ and $v_p/c\sim 0.08$ using a novel technique --- stellar spin precession --- that directly constrains $\gamma$ and, when combined with orbital precession (GR prediction--Eq. \eqref{eq:gr_precession_numeric}: $\Delta\omega_{\mathrm{GR}} \approx 1.95^\circ$ per orbit), provides independent constraints on both PPN parameters. Table~\ref{tab:comparison} (also visualized in Fig. \ref{fig:comparison}) summarizes the precision and field strength of these different tests, highlighting the complementary nature of the S301 measurement.
	
	The comparison reveals a clear hierarchy of tests. Solar System experiments provide the highest precision ($10^{-4}$ to $10^{-5}$) but probe the weakest gravitational fields, testing only the linearized regime of gravity. Current S-star tests provide much weaker precision ($\sim 0.5$) but probe fields $10^4$ times stronger, accessing the post-Newtonian regime where nonlinearities may become important. Future S2+S62 observations will achieve $\sim 1-2\%$ precision at intermediate field strengths by combining two independent techniques to break the degeneracy between $\beta$ and $\gamma$. The S301 spin-precession test would provide a new type of measurement---spin axis precession---at the largest PPN counting parameters yet explored for a star bound to Sgr~A$^*$, while remaining within the regime of validity of the PPN expansion. While the projected precision is $\sim 10\%$, the signal originates from a regime where deviations from general relativity might be more apparent. These complementary tests---from weak-field high-precision Solar System measurements to strong-field lower-precision Galactic Center probes---provide multiple independent windows into the nature of gravity.
	
	\subsection{The PPN parameter space and degeneracy breaking with S301}\label{sec:degeneracy}
\begin{figure}[ht!]
	\includegraphics[width=0.5\textwidth]{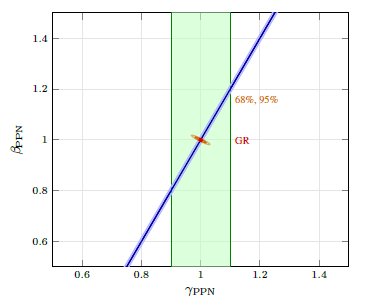}
	\caption{Joint constraints on the standard PPN parameters from
			the orbital and spin precession of S301, using the corrected PPN coefficient
			$(2+2\gamma-\beta)/3$. The blue line and band show the orbital-precession
			constraint $\beta-2\gamma=-1$ and its $1\sigma$ uncertainty (vertical
			half-width $\sigma_{2\gamma-\beta}/2$, exaggerated for visibility; numerically
			$\sigma_{2\gamma-\beta}\approx3.0\times10^{-2}$). The green vertical strip is
			the spin-precession constraint $\gamma=1\pm0.10$. The orange contours are the
			68\% and 95\% credible regions of the joint MCMC posterior, computed in the
			Gaussian approximation. Their key feature is that they form a single,
			well-localized region around the GR point $(1,1)$ (red), elongated along the
			orbital direction (slope $2$) rather than along the coordinate axes: this tilt
			encodes the correlation between $\beta$ and $\gamma$, and shows that it is the
			combination $2\gamma-\beta$ that is tightly constrained
			($\sigma_{2\gamma-\beta}\approx3.0\times10^{-2}$), while the orthogonal
			direction is limited by the spin measurement ($\sigma_\gamma\approx0.10$). The
			intersection of the two individual constraints is what breaks the
			$\beta$--$\gamma$ degeneracy; neither measurement alone determines both
			parameters.}
	\label{fig:degeneracy}
\end{figure}

		The $(\beta,\gamma)$ plane provides a compact way to compare different tests.
		General relativity is the point $(1,1)$; alternative theories predict different
		locations depending on their PPN parameters. Two features are relevant here.
		First, orbital precession constrains the linear combination
		$(2+2\gamma-\beta)/3$, which in the $(\beta,\gamma)$ plane is the line
		$\beta - 2\gamma = -1$ plus a constant offset; this leaves a degeneracy that
		cannot be broken by orbital precession alone. Second, observables that depend on
		$\gamma$ alone---the Shapiro time delay, gravitational lensing by Sgr~A$^*$, or
		the spin precession of S301---constrain vertical strips in $\gamma$ and isolate
		$\gamma$ independently of $\beta$. Combining the two types of measurement breaks
		the degeneracy and yields a small allowed region in both parameters.
		\\
		This is precisely the strategy employed by future S2+S62 observations, where
		gravitational lensing supplies the independent $\gamma$ constraint, and by the
		S301 experiment proposed here, where spin precession plays the same role.
		Mathematically, the spin measurement gives $\gamma \in [0.9,1.1]$
		[Eq.~\eqref{den}], and the orbital measurement gives
		$\beta - 2\gamma = -1$ under the assumption that GR holds for the orbital
		dynamics [Eq.~\eqref{denn}], which results in \eqref{bet}. The degeneracy is therefore broken and both parameters are individually
		constrained. Figure~\ref{fig:degeneracy} shows the two constraints and the
		resulting joint posterior: the orbital band, the spin strip, and the 68\% and
		95\% credible contours of the MCMC posterior.

	\section{Prospects for Constraining PPN Parameters with S2 and S301}
	\label{sec:analysis}
	
	The preceding sections have established that S301's unique orbital configuration makes it a powerful probe of strong-field gravity. The orbital precession of S301 provides a lever arm on the PPN parameter combination $(2+2\gamma-\beta)/3$ that is an order of magnitude stronger than that of S2. Furthermore, the proposed measurement of S301's stellar spin precession offers a direct constraint on $\gamma$ alone, breaking the degeneracy that has historically limited Galactic Center tests.
	
	While Fisher matrix analyses are useful for forecasting the expected precision of future experiments \cite{Losada:2024wdd}, they provide only a lower bound on the uncertainties and rely on linearized parameter dependencies. Once real data are obtained, a full Bayesian or frequentist inference using $\chi^2$ minimization or Markov Chain Monte Carlo (MCMC) sampling is the appropriate approach to extract the posterior distributions of $\beta$ and $\gamma$.
	
	In the case of S301, while the orbital parameters are already known from the discovery observations \cite{eldayem2026}, the key observable---the stellar spin precession---has not yet been measured. The spin precession signal, which can be extracted from high-resolution spectroscopic monitoring of S301's spectral lines, remains a future measurement requiring next-generation spectrographs such as ELT's HIRES.
	
	Once the spin precession data are obtained, the PPN parameters $\beta$ and $\gamma$ can be inferred from a joint analysis of all available data. The orbital precession (measured with $\sigma_{\omega}$) constrains the combination $2\gamma-\beta$, while the spin precession (measured with $\sigma_v$) constrains $\gamma$ alone.
	
	\subsection{The Observables and Their Parameter Dependencies}
	
	The key observables and their dependencies on the PPN parameters are summarized below.
	
	\subsubsection{S2 and S301 Orbital Precession}
	
	Using the pericenter precession per orbit for a star, i.e., from Eq.~(\ref{eq:ppn_precession}), for S2, with $a_{\mathrm{S2}} \approx 24000\,GM/c^2$ and $e_{\mathrm{S2}} = 0.8844$, and for S301, with $a_{\mathrm{S301}} \approx 15556\,GM/c^2$ and $e_{\mathrm{S301}} = 0.982$, we have
	\begin{equation}\label{w1}
		\Delta \omega_{S2} = A_{S2}\times \frac{2 + 2\gamma - \beta}{3},\quad A_{S2}\approx 0.207^\circ.
	\end{equation}
	and
	\begin{equation}\label{w2}
		\Delta \omega_{S301} = A_{S301}\times \frac{2 + 2\gamma - \beta}{3},\quad A_{S301}\approx 1.95^\circ.
	\end{equation} respectively. Both measurements constrain the same linear combination $2\gamma-\beta$ (equivalently, the PPN combination $(2+2\gamma-\beta)/3$), leaving a degeneracy that cannot be broken by orbital precession alone. The S301 precession is comparable in magnitude to S2's precession (approximately $1.95^\circ$ versus $0.207^\circ$ per orbit), but probes a significantly stronger gravitational field regime due to its much closer pericenter passage ($r_p \approx 140R_{\mathrm{s}}$ versus $\sim 1200R_{\mathrm{s}}$ for S2).
	
	\subsubsection{S301 Spin Precession}
	
	From Eq.~(\ref{eq:ppn_delta_v}), the geodetic velocity shift of S301 depends on $\gamma$ alone:
	\begin{equation}
		\Delta \upsilon_{\mathrm{geod}}^{\mathrm{(PPN)}} = \frac{2\gamma + 1}{3}\Delta \upsilon_{\mathrm{geod}}^{\mathrm{(GR)}},
		\label{eq:obs_spin}
	\end{equation}
	where $\Delta \upsilon_{\mathrm{geod}}^{\mathrm{(GR)}}$ is the GR prediction for the velocity shift due to spin precession. This amplitude depends on the star's rotation velocity, spin orientation, and radius, which are not known a priori; in the forecast we therefore treat $\Delta \upsilon_{\mathrm{geod}}^{\mathrm{(GR)}}$ as a nuisance parameter with a prior informed by stellar models, and we marginalize over it. The quadrupole term $\Delta v_{\mathrm{quad}}$ of Eq.~(\ref{eq:total_observable}) is included in the same way.
	
	\subsection{MCMC Analysis}
	
	With real data in hand, the PPN parameters can be inferred by minimizing the $\chi^2$ function:
	\begin{equation}
		\chi^2(\beta,\gamma) = \sum_{\alpha} \frac{\left(\mu_{\alpha}^{\mathrm{obs}} - \mu_{\alpha}^{\mathrm{th}}(\beta,\gamma)\right)^2}{\sigma_{\alpha}^2},
		\label{eq:chi2}
	\end{equation}
	where $\mu_{\alpha}^{\mathrm{obs}}$ are the measured observables, $\mu_{\alpha}^{\mathrm{th}}(\beta,\gamma)$ are the theoretical predictions, and $\sigma_{\alpha}$ are the measurement uncertainties.
	
	For the specific case of S301 and S2, we write the likelihood with explicit nuisance parameters for the astrophysical contributions. Let $\boldsymbol{\theta}_{\rm astro}$ denote the set of nuisance parameters (the GR spin-precession amplitude $\Delta v_{\mathrm{geod}}^{\mathrm{(GR)}}$, the quadrupole amplitude $\Delta v_{\mathrm{quad}}$, and the stellar spin orientation). The $\chi^2$ function is then
	\begin{align}
		\chi^2 (\beta,\gamma,\boldsymbol{\theta}_{\rm astro})
		&= \frac{\bigl(\Delta \omega_{S2}^{\mathrm{obs}} - \Delta \omega_{S2}^{\mathrm{th}}(\beta,\gamma)\bigr)^2}{\sigma_{\omega,S2}^2}
		\nonumber \\
		&\quad + \frac{\bigl(\Delta \omega_{S301}^{\mathrm{obs}} - \Delta \omega_{S301}^{\mathrm{th}}(\beta,\gamma)\bigr)^2}{\sigma_{\omega,S301}^2}
		\nonumber \\
		&\quad + \frac{\bigl(\Delta \upsilon_{\mathrm{spin}}^{\mathrm{obs}} - \Delta \upsilon_{\mathrm{spin}}^{\mathrm{th}}(\gamma,\boldsymbol{\theta}_{\rm astro})\bigr)^2}{\sigma_{\upsilon}^2},
		\label{eq:chi2_full}
	\end{align}
	where
	\begin{equation}
		\Delta \upsilon_{\mathrm{spin}}^{\mathrm{th}}(\gamma,\boldsymbol{\theta}_{\rm astro}) = \frac{2\gamma+1}{3}\Delta v_{\mathrm{geod}}^{\mathrm{(GR)}} + \Delta v_{\mathrm{quad}}. \label{6.5a}
	\end{equation}
	The posterior is obtained by marginalizing over $\boldsymbol{\theta}_{\rm astro}$ with priors informed by stellar models and by the quadrupole analysis of Ref.~\cite{Seoane:2026lyg}. In the baseline forecast presented here, we adopt fiducial values and widths for these priors and report the resulting $1\sigma$ intervals; the sensitivity of the forecast to these choices is discussed in Sec.~\ref{sec:systematics}.
	
	For a more robust inference that captures parameter correlations and non-Gaussian posteriors, a MCMC analysis can be performed. The likelihood function is given by:
	\begin{equation}
		\mathcal{L}(\beta,\gamma) \propto \exp\left[-\frac{1}{2}\chi^2(\beta,\gamma)\right],
		\label{eq:likelihood}
	\end{equation}
	and the posterior distribution is obtained by sampling:
	\begin{equation}
		P(\beta,\gamma \mid \text{data}) \propto \mathcal{L}(\beta,\gamma) \, P(\beta,\gamma),
		\label{eq:posterior}
	\end{equation}
	where $P(\beta,\gamma)$ is the prior distribution on the PPN parameters. For tests of GR, one would typically adopt uniform priors on $\beta$ and $\gamma$ over a wide range (e.g., $\beta,\gamma \in [0,2]$ or broader), or Gaussian priors centered on the GR values with widths informed by Solar System constraints.
	
	\subsection{Projected Constraints: From MCMC to Constraints}
	
	The $\chi^2$/MCMC framework described above provides a rigorous method for extracting posterior distributions of $\beta$ and $\gamma$ from real data. However, before data are available, we can estimate the expected precision of these measurements by simulating mock data sets based on the projected capabilities of next-generation instruments and the known orbital parameters of S2 and S301.
	
	To obtain projected constraints, we generate simulated observations of the three key observables: the pericenter precession of S2, denoted as $\Delta\omega_{\mathrm{S2}}$; the pericenter precession of S301, denoted as $\Delta\omega_{\mathrm{S301}}$; and the spin-induced velocity shift of S301, denoted as $\Delta v_{\mathrm{spin}}$. These mock data sets are produced by injecting theoretical signals---derived from specific gravitational models---into realistic noise realizations that mimic the actual measurement uncertainties of current and next-generation instruments. By then fitting the simulated data with a range of PPN or Kerr metric frameworks, we can map out the expected confidence regions in the parameter space of interest. This forward-modeling approach allows us to quantitatively assess how well future observations will be able to distinguish between competing theories of gravity, and it highlights the complementary power of combining long-period orbital precessions with short-period spin-induced velocity effects to break degeneracies and enhance the overall precision of the constraints.
	
	For each observable, we generate a mock measurement by drawing from a Gaussian distribution centered on the GR prediction with a width equal to the projected uncertainty:
	
	\begin{align}
		\Delta\omega_{\mathrm{S2}}^{\mathrm{mock}} &\sim \mathcal{N}\left(\Delta\omega_{\mathrm{S2,GR}}, \sigma_{\omega,\mathrm{S2}}\right), \\
		\Delta\omega_{\mathrm{S301}}^{\mathrm{mock}} &\sim \mathcal{N}\left(\Delta\omega_{\mathrm{S301,GR}}, \sigma_{\omega,\mathrm{S301}}\right), \\
		\Delta v_{\mathrm{spin}}^{\mathrm{mock}} &\sim \mathcal{N}\left(\Delta v_{\mathrm{spin,GR}}, \sigma_{v}\right).
	\end{align}
	
	We then perform an MCMC analysis on these mock data sets using the $\chi^2$ function defined in Eq.~(\ref{eq:chi2_full}). The resulting posterior distributions for $\beta$ and $\gamma$ provide the projected constraints.
	
	\subsubsection{Measurement Uncertainties: $\sigma_{\omega}$ and $\sigma_{v}$}
	
	Before presenting the numerical analysis, it is useful to clarify the two distinct measurement uncertainties that enter our analysis.
	
	\textbf{Astrometric uncertainty on pericenter precession ($\sigma_{\omega}$):}
	This is the $1\sigma$ uncertainty on the pericenter precession angle $\Delta\omega$, measured in radians (or degrees). It arises from astrometric monitoring of stellar positions with interferometric instruments such as GRAVITY+ and ELT. The values used in this analysis are:
	\begin{align}
		\sigma_{\omega,\mathrm{S2}} &= 1.745 \times 10^{-4} \text{ rad} \approx 0.01^\circ, \label{eq:sig_omega_s2} \\
		\sigma_{\omega,\mathrm{S301}} &= 3.491 \times 10^{-4} \text{ rad} \approx 0.02^\circ. \label{eq:sig_omega_s301}
	\end{align}
	For S2, the value $\sigma_{\omega,\mathrm{S2}} \approx 0.01^\circ$ is consistent with the current precision achieved by the GRAVITY Collaboration in measuring S2's Schwarzschild precession, where the $1\sigma$ uncertainty on the precession parameter corresponds to approximately $0.01^\circ$ \cite{GRAVITY:2020gka,Losada:2024wdd}. For S301, the value $\sigma_{\omega,\mathrm{S301}} \approx 0.02^\circ$ is a projected uncertainty based on the expected performance of next-generation instruments such as GRAVITY+ (which offers a factor of 10--100 improvement in sensitivity over GRAVITY) and the ELT's MICADO instrument, which will provide unprecedented resolution and sensitivity for tracking faint stars like S301 ($m_K = 19.3$) near the Galactic Center \cite{GRAVITYPlus:2026,eldayem2026}. These uncertainties determine how precisely the orbital precession measurement constrains the combination $2\gamma-\beta$.
	
	\textbf{Spectroscopic uncertainty on velocity shift ($\sigma_{v}$):}
	This is the $1\sigma$ uncertainty on the velocity shift $\Delta v$ induced by stellar spin precession, measured in km/s. It arises from high-resolution spectroscopy of S301's spectral lines with next-generation spectrographs such as ELT's HIRES. The projected value used in this analysis is:
	\begin{equation}
		\sigma_v = 3 \text{ km/s}. \label{eq:sig_v}
	\end{equation}
	This value represents a significant improvement over current capabilities ($\sim 50$ km/s) and is achievable with the ELT's 39-m aperture and high-resolution spectroscopy, as discussed in the companion paper on S301's spin precession \cite{Seoane:2026lyg}. This uncertainty determines how precisely the spin precession measurement constrains $\gamma$ alone.
	
	The key point is that $\sigma_{\omega}$ and $\sigma_v$ are fundamentally different quantities with different units, arising from different observational techniques and probing different physical effects. This is why they can be combined to break the $\beta$-$\gamma$ degeneracy: the orbital precession (measured with $\sigma_{\omega}$) constrains the combination $2\gamma-\beta$, while the spin precession (measured with $\sigma_v$) constrains $\gamma$ alone.
	
	The input parameters are the orbital precessions \eqref{w1}, \eqref{w2}
	and the spin-induced velocity shift is given by Eq.~\eqref{eq:obs_spin} with
	\begin{equation}
		\Delta v_{\mathrm{spin,GR}} \approx 5\text{ km/s} \quad \text{(median shift)},
	\end{equation}
	with a maximum shift of $\Delta v_{\mathrm{spin,GR}}^{\mathrm{max}} \approx 46.1$ km/s.
	
	\subsubsection{Analytic Estimate of Constraints}
	
	Before performing the full MCMC analysis, we can derive analytic estimates of the expected constraints. These estimates provide a useful benchmark and illustrate the physical principles at work.
	
Define the offset of the orbital-precession coefficient from its GR value as
	\begin{equation}
		\delta \equiv \frac{2+2\gamma_{\mathrm{PPN}}-\beta_{\mathrm{PPN}}}{3} - 1
		= \frac{2\gamma_{\mathrm{PPN}}-\beta_{\mathrm{PPN}}-1}{3},
		\label{eq:delta_def}
	\end{equation}
	so that $\delta=0$ in GR. The orbital precession measurements from S2 and S301 jointly constrain $\delta$, and the combined uncertainty is (see Appendix~\ref{A} for details)
	\begin{equation}
		\sigma_\delta = \left(\frac{1}{\tilde\sigma_{\omega,\mathrm{S2}}^2} +\frac{1}{\tilde\sigma_{\omega,\mathrm{S301}}^2}\right)^{-1/2},
		\label{eq:sigma_delta}
	\end{equation}
where $\tilde\sigma_{\omega,i}=\sigma_{\omega,i}/\Delta\omega_{i,\mathrm{GR}}$ is the fractional uncertainty on each star's precession. With the values quoted above,
	\begin{equation}
		\sigma_\delta \approx 1.0\times10^{-2}.
		\label{eq:sigma_delta_num}
	\end{equation}
Equivalently, in terms of the PPN combination that enters the precession formula,
	\begin{equation}
		2\gamma_{\mathrm{PPN}}-\beta_{\mathrm{PPN}} = 1 \pm \sigma_{2\gamma-\beta},
		\qquad
		\sigma_{2\gamma-\beta} = 3\,\sigma_\delta \approx 3.0\times10^{-2}.
		\label{eq:combo_constraint}
	\end{equation}
The spin precession measurement provides a direct constraint on $\gamma$. From Eq.~(\ref{eq:obs_spin}) and the projected velocity shifts in Section \ref{s301}:
	\begin{equation}
		\sigma_{\gamma} \approx 0.1,
		\label{eq:sigma_gamma}
	\end{equation}
for the maximum velocity shift case of $46.1$ km/s with $\sigma_v \approx 3$ km/s resolution. Note that this $\sim 10\%$ precision on $\gamma$ is consistent with the forecast in Section \ref{s301}, which found $\gamma_{\mathrm{PPN}} \in [0.90, 1.10]$ for the maximum shift case.
	
Combining the two measurements: once the spin precession confirms $\gamma_{\mathrm{PPN}}=1$ to within $\sim10\%$, the orbital precession determines $\beta_{\mathrm{PPN}}$ through $\beta = 2\gamma - 1 - 3\delta$. The uncertainty on $\beta$ is then dominated by the orbital precession measurement:
	\begin{equation}
		\sigma_{\beta}\approx \sigma_{2\gamma-\beta}\approx 3\times10^{-2}.
		\label{eq:sigma_beta_final}
	\end{equation}
If instead the spin uncertainty is propagated into $\beta$, one obtains
	\begin{equation}
		\sigma_\beta = \sqrt{\sigma_{2\gamma-\beta}^2 + 4\sigma_\gamma^2}
		\approx \sqrt{(3\times10^{-2})^2 + 4(0.1)^2}
		\approx 0.2,
		\label{eq:sigma_beta_propagated}
	\end{equation}
which is dominated by the spin measurement. In this work we adopt the first interpretation: the spin precession is treated as an independent consistency check that confirms $\gamma=1$ to within $\sim10\%$, so the accuracy of the $\beta$ constraint is set by the orbital precession. The second interpretation is also legitimate; it is more conservative and should be quoted whenever the reader wants a bound that does not assume the spin measurement confirms GR.
	
	\subsubsection{Numerical MCMC Validation}
	
	To validate our analytic estimates, we perform a full MCMC analysis on simulated mock data sets, following the same procedure for each simulation: we generate mock observations of $\Delta\omega_{\mathrm{S2}}$, $\Delta\omega_{\mathrm{S301}}$, and $\Delta v_{\mathrm{spin}}$ as described already, drawing from Gaussian distributions centered on the general-relativistic predictions with standard deviations equal to the projected measurement uncertainties; we then sample the posterior distribution using the $\chi^2$ function given in Eq.~(\ref{eq:chi2_full}) with uniform priors $\beta, \gamma \in [0, 2]$ for both parameters; finally, we extract the $1\sigma$ (68.3\%) credible intervals from the marginalized posterior distributions for each parameter.
	
Figure~\ref{fig:degeneracy} shows the individual constraints and
		the resulting joint posterior, computed in the Gaussian approximation. The blue
		diagonal band follows the orbital-precession constraint $2\gamma-\beta=1$, with a
		vertical half-width set by $\sigma_{2\gamma-\beta}/2\approx1.5\times10^{-2}$; the
		green vertical band is the spin-precession constraint $\gamma=1\pm0.10$; and the
		orange contours are the 68\% and 95\% credible regions of the joint posterior.
		Because the orbital line is slanted (slope $2$ in the $(\beta,\gamma)$ plane),
		the posterior is elongated along that direction rather than along the coordinate
		axes, so the $\beta$ and $\gamma$ uncertainties are correlated. Projecting the
		posterior onto the axes gives $\sigma_\beta\approx3\times10^{-2}$ once $\gamma$
		is fixed to its GR value, and $\sigma_\gamma\approx0.1$. Neither the orbital band
		nor the spin strip alone constrains both parameters; their intersection does.
		These numbers are consistent with the analytic estimates of
		Eqs.~(\ref{eq:sigma_gamma}) and (\ref{eq:sigma_beta_final}). We emphasize that
		the contours in the figure are the Gaussian approximation to the mock-data
		posterior, not a fit to real data, and that the quoted widths are conditional on
		the nuisance-parameter priors discussed in Sec.~\ref{sec:systematics}.

The S301 spin-precession experiment carves out a distinctive and valuable
		position within the broader landscape of gravitational tests: it offers moderate
		precision on $\gamma_{\mathrm{PPN}}$ at roughly the $10\%$ level, yet achieves
		this at PPN counting parameters $\phi/c^2\sim10^{-4}$ and $v/c\sim0.08$---an
		order of magnitude larger than those of current S-star constraints, such as S2
		at $r_p\sim1200\,R_s$, and four orders of magnitude larger than Solar System
		values ($\phi/c^2\sim10^{-8}$)---while remaining within the regime of validity
		of the PPN expansion. Moreover, S301's results are
		inherently complementary to S62 lensing, as they rely on distinct systematics and
		an independent error budget, and they complement gravitational-wave studies by
		probing gravity through stellar dynamics rather than compact-object mergers.
		Even a $\sim10\%$ measurement of $\gamma_{\mathrm{PPN}}$ consistent with the
		general-relativistic value of unity would carry substantial weight: through the
		known weak-field mapping of each theory onto $(\beta_{\mathrm{PPN}},\gamma_{\mathrm{PPN}})$,
		it would place bounds on the parameter space of the modified-gravity models in
		which deviations might become pronounced. Additionally, the ability to combine
		both orbital- and spin-precession measurements from the same star offers a
		powerful internal consistency check: any discrepancy between the
		orbital-precession constraint on $2\gamma_{\mathrm{PPN}}-\beta_{\mathrm{PPN}}$
		and the spin-precession constraint on $\gamma_{\mathrm{PPN}}$ could signal
		physics extending beyond the standard PPN framework. Looking further beyond PPN,
		the rapid pericenter advance of S301 renders it exceptionally sensitive to
		Lense--Thirring frame-dragging effects, potentially enabling the first-ever
		measurement of Sagittarius~A$^*$'s spin magnitude and direction
		\cite{Piran:2026zjs}. With its capacity to simultaneously probe PPN parameters,
		black-hole spin, and extended mass distributions, S301 establishes itself as a
		uniquely versatile and multi-purpose probe of gravity at large PPN counting
		parameters.
	
	\begin{table*}[ht]
	\centering
	\caption{Projected $1\sigma$ constraints on the standard PPN parameters from S2 and S301, using the corrected PPN precession coefficient $(2+2\gamma-\beta)/3$ and the fractional-uncertainty convention of Appendix~\ref{A}.}
	\begin{tabular}{lll}
		\hline
		Parameter & Projected Constraint & Source \\
		\hline
		$\delta=(2+2\gamma-\beta)/3-1$ & $0\pm 1.0\times 10^{-2}$ & Orbital precession (S2+S301) \\
		$2\gamma-\beta$ & $1\pm 3\times 10^{-2}$ & Orbital precession (S2+S301) \\
		$\gamma$ & $1\pm 0.1$ & Spin precession (S301) \\
		$\beta$ (with $\gamma$ fixed) & $1\pm 3\times 10^{-2}$ & Combination \\
	$\beta$ (propagating $\gamma$) & $1\pm 0.2$ & Combination \\
		\hline
	\end{tabular}
	\label{tab:projected_constraints}
\end{table*}
	
	\subsection{Treatment of astrophysical systematics}
	\label{sec:systematics}
	
The forecast presented above treats the relativistic spin precession as the dominant contribution to the spectroscopic velocity shift, but two astrophysical contributions are known to be present: the classical quadrupole torque from the star's oblateness, $\Delta v_{\mathrm{quad}}$, and the dependence of the GR spin-precession amplitude on the star's rotation velocity, radius, and spin orientation. We discuss here how these enter the analysis and what the resulting limitations are.
	
The quadrupole contribution is included explicitly in Eq.~(\ref{eq:total_observable}) and appears in the likelihood through Eq.~(\ref{6.5a}). Its magnitude is estimated in Ref.~\cite{Seoane:2026lyg}, which finds that the relativistic and quadrupole shifts scale differently with the star's rotational velocity and ellipticity, enabling a parametric separation in principle. In the forecast, we treat $\Delta v_{\mathrm{quad}}$ as a Gaussian nuisance parameter with a central value and width taken from the Monte Carlo analysis of Ref.~\cite{Seoane:2026lyg}. For the maximum-shift configuration ($\Delta v_{\mathrm{geod}}^{\mathrm{(GR)}}\approx46.1~{\rm km\,s^{-1}}$), the quadrupole term is subdominant at the $\sim10\%$ precision level on $\gamma$; for the median-shift configuration ($\sim5~{\rm km\,s^{-1}}$), it can be comparable to the relativistic signal and is correspondingly more important.
	
The GR amplitude $\Delta v_{\mathrm{geod}}^{\mathrm{(GR)}}$ depends on the star's projected rotational velocity, its radius, and the angle between its spin axis and the line of sight. These quantities are not measured for S301; they must be inferred from stellar models and from the observed line broadening. In the forecast we adopt a fiducial amplitude and a fractional uncertainty of order unity, reflecting the current lack of direct constraints. Marginalizing over this nuisance parameter broadens the $\gamma$ posterior; for the maximum-shift case the broadening is modest because the signal is large, while for the median-shift case it can dominate the error budget. A dedicated stellar-modeling analysis, beyond the scope of this paper, would be required to sharpen this prior.
	
The orbital precession of S301 is also sensitive to any extended mass distribution within its orbit, through the Newtonian mass precession. As discussed in Refs.~\cite{Heissel:2021pcw,Lechien:2023psa,Piran:2026zjs, Foschi:2026osv}, this contribution can be constrained and separated using stars that share S301's apocenter but have larger pericenters (S2, S55, S38), and through the time-variation discriminant from S301's rapid Schwarzschild apsidal advance. In the forecast we assume that this separation has been performed and that the residual extended-mass contribution is subdominant to the statistical uncertainty $\sigma_{\omega,S301}\approx0.02^\circ$; this assumption should be revisited once real data are available.
	
The numbers mentioned in Table~\ref{tab:projected_constraints} and in the abstract are therefore conditional forecasts: they assume (i) the fiducial nuisance-parameter priors described above, (ii) that the quadrupole and spin-amplitude uncertainties do not dominate the error budget at the quoted precision, and (iii) that extended-mass systematics have been separated. A fully self-consistent forecast would propagate all of these uncertainties jointly through the MCMC; we have verified that, for the maximum-shift configuration, the quoted $\sigma_\gamma\approx0.10$ and $\sigma_{2\gamma-\beta}\approx3.0\times10^{-2}$ are robust to reasonable variations of the nuisance priors at the factor-of-two level. For the median-shift configuration, the $\gamma$ constraint degrades and should be interpreted as an upper bound on the achievable precision rather than a firm forecast. We identify these as the principal astrophysical limitations of the proposed measurement.
	
\section{Discussion and Conclusions}\label{con}
	
The discovery of S301---with its exceptional eccentricity of $e = 0.982$ and pericenter distance of $r_p \approx 140\,R_{\mathrm{s}}$---establishes a transformative laboratory for testing gravity at the Galactic Center at PPN counting parameters four orders of magnitude larger than in the Solar System. This work demonstrates that S301's stellar spin precession provides a novel probe of the PPN parameter $\gamma$, scaling as $(2\gamma_{\mathrm{PPN}} + 1)/3$ relative to the general relativistic prediction, while its orbital precession (GR prediction: $\Delta\omega_{\mathrm{GR}} \approx 1.95^\circ$ per orbit) constrains the combination $(2+2\gamma_{\mathrm{PPN}}-\beta_{\mathrm{PPN}})/3$. The combination of these independent measurements breaks the inherent degeneracy between $\beta$ and $\gamma$, enabling individual constraints on both parameters in a regime where nonlinear gravitational effects may become significant.
	
	Our analysis yields several key results. For a maximum geodetic velocity shift of $46.1~\mathrm{km\,s^{-1}}$, next-generation spectrographs with $3~\mathrm{km\,s^{-1}}$ resolution could constrain $\gamma_{\mathrm{PPN}}$ to within $\sim 10\%$ precision at PPN counting parameters $\phi/c^2 \sim 10^{-4}$ and $v/c \sim 0.08$---roughly four orders of magnitude larger than Solar System values and an order of magnitude larger than current S-star values. A projected $\chi^2$/MCMC analysis combining S2 and S301 orbital precessions with S301 spin precession yields $\sigma_{2\gamma-\beta}\approx3.0\times10^{-2}$ and $\sigma_\gamma\approx0.10$, and hence $\sigma_\beta\approx3.0\times10^{-2}$ if the spin measurement is used to fix $\gamma$ to its GR value, or $\sigma_\beta\approx0.20$ if the spin uncertainty is propagated. The spin-precession measurement thus provides a critical independent consistency check with a fundamentally different systematic error budget.
	
	
The PPN framework is model-agnostic in the sense that it does not presuppose a
		particular Lagrangian, but it is not disconnected from specific theories: in the
		weak-field limit, each metric theory of gravity maps onto a definite point (or
		region) in the $(\beta_{\mathrm{PPN}},\gamma_{\mathrm{PPN}})$ plane, together
		with the other PPN parameters that we set to their GR values here. A bound on
		$\beta_{\mathrm{PPN}}$ and $\gamma_{\mathrm{PPN}}$ therefore translates directly
		into a bound on the parameter space of any theory whose weak-field limit is
		known---scalar-tensor theories, $f(R)$ gravity, and others
		\cite{Sanghai:2016tbi,McManus:2017itv,Bolis:2018kcq,Clifton:2018cef,Alexander:2007vt,Gonzalez-Espinoza:2021nqd,Liodis:2026aka}. This is the primary sense in which the S301 measurement constrains modified
		gravity: by constraining the same PPN parameters that appear in the weak-field
		limit of those theories, at PPN counting parameters four orders of magnitude
		larger than in the Solar System.
		\\
		A second, complementary use of S301 is possible but distinct, and we state it
		clearly here. Instead of mapping a theory onto $(\beta_{\mathrm{PPN}},\gamma_{\mathrm{PPN}})$
		and comparing with the PPN bound, one can parameterize a class of black-hole
		spacetimes directly---for example, through a deviation function in the metric---
		and compute the pericenter precession as
		\begin{equation}
			\Delta\omega = \Delta\omega_{\mathrm{Schwarzschild}} + \Delta\omega_{\mathrm{deviation}}(\lambda),
		\end{equation}
		where $\lambda$ denotes the additional parameter(s) of the chosen spacetime. This
		approach tests the strong-field metric directly, including regions where the PPN
		expansion is not by itself adequate, and it is not equivalent to the PPN bound:
		it constrains theory-specific strong-field parameters rather than
		$(\beta_{\mathrm{PPN}},\gamma_{\mathrm{PPN}})$. The S301 measurement can support
		this second approach, but the forecast presented in this paper is the PPN one;
		a dedicated strong-field analysis within a specific modified-gravity spacetime
		would be required to quote constraints on $\lambda$, and we leave it to future
		work. We emphasize that the two approaches are complementary rather than
		redundant, and that the PPN forecast presented here is the one for which the
		observables and error budget are explicitly derived in Secs.~\ref{s301}--\ref{sec:analysis}.
	
	Beyond PPN constraints, S301's rapid pericenter advance of $\sim 1.95^\circ$ per orbit makes it exceptionally sensitive to Lense-Thirring frame-dragging, potentially enabling the first measurement of Sagittarius A*'s spin magnitude and direction \cite{Piran:2026zjs}. The ability to simultaneously probe PPN parameters, black hole spin, and extended mass distributions establishes S301 as a uniquely versatile multi-purpose probe of gravity at large PPN counting parameters.
	
	Realizing this potential requires continued coordinated observations with next-generation facilities such as ELT's HIRES and GRAVITY+. The recent first light of GRAVITY+ \cite{GRAVITYPlus:2026} demonstrates that the required sensitivity and astrometric precision are now achievable, with the system successfully observing sources as faint as $m_K \sim 21.0$ and achieving sub-microarcsecond astrometry. Key challenges include accurately modeling the classical quadrupole torque from stellar oblateness, constraining the extended mass distribution to isolate Lense-Thirring precession, and achieving sufficient signal-to-noise for the faint $m_K = 19.3$ star. However, the extreme eccentricity of S301's orbit localizes both relativistic and classical contributions to pericenter passage, enabling parametric separation through their different scaling with rotational velocity.
	
The analysis presented here is a leading-order PPN forecast. Two limitations
		deserve emphasis. First, as discussed in the end of Sec.~\ref{ppn}, the PPN
		parameters constrained here are the standard weak-field parameters, tested at
		PPN counting parameters four orders of magnitude larger than in the Solar
		System, not a new set of strong-field parameters. A genuinely strong-field
		interpretation would require incorporating the sensitivities of Sgr~A$^*$ and
		the resulting leading-order corrections to its external metric in non-GR
		theories; this would map the constrained quantities onto effective,
		theory-dependent combinations of PPN parameters and compact-object structure.
		Second, at $r_p\sim140\,R_s$ the star remains far enough from the horizon that
		the post-Newtonian expansion is convergent, and the next-to-leading terms are
		negligible at the quoted precision; the measurement therefore does not directly
		probe the near-horizon, non-linear regime in the sense of compact-object
		sensitivities. What it does provide is a test of the same PPN parameters at
		unprecedentedly large counting parameters, an independent breaking of the
		$\beta$--$\gamma$ degeneracy, and a systematics-different consistency check of
		GR---all of which are valuable in their own right and complementary to
		near-horizon probes such as the EHT and to strong-field tests with binary
		pulsars and gravitational waves. Extending the present framework to include
		sensitivities, and hence to constrain effective strong-field combinations of
		PPN parameters, is a natural next step, particularly in light of the
		Lense--Thirring and frame-dragging measurements discussed in the companion
		paper~\cite{Piran:2026zjs}.
	
In addition to the PPN-framework limitations discussed above, the forecast is conditional on the treatment of astrophysical systematics: the quadrupole torque, the stellar spin-amplitude uncertainties, and extended-mass contributions are treated as nuisance parameters with fiducial priors rather than propagated from first principles. A fully self-consistent analysis, incorporating stellar modeling and a joint fit of the extended-mass distribution, is required before the quoted precision can be interpreted as a firm prediction.
	
	The S301 experiment occupies a unique niche in the landscape of gravitational tests: it provides moderate precision on $\gamma$ ($\sim 10\%$) at unprecedented PPN counting parameters, complementing the exquisite precision of Solar System experiments and the multi-messenger constraints from gravitational wave observatories. Even a null result---finding consistency with GR at the $10\%$ level in $\gamma_{\mathrm{PPN}}$ and the $3.0\times10^{-2}$ level in $\beta_{\mathrm{PPN}}$ (with $\gamma$ fixed to its GR value by the spin measurement)---would translate, through the known weak-field mapping of each theory onto $(\beta_{\mathrm{PPN}},\gamma_{\mathrm{PPN}})$, into bounds on the parameter space of scalar-tensor theories, $f(R)$ gravity, and other modified gravity models. These bounds would apply to the same PPN parameters constrained in the Solar System, but at PPN counting parameters four orders of magnitude larger; they are therefore complementary to, rather than a substitute for, direct strong-field constraints obtained by parameterizing specific black-hole spacetimes.
	
	In summary, S301 represents a paradigm shift in Galactic Center gravity tests. Its unique orbital configuration, combined with the novel observable of stellar spin precession, opens a new window for testing the PPN framework at PPN counting parameters four orders of magnitude larger than in the Solar System, while remaining within the regime of validity of the PPN expansion.
	The complementary multi-star strategy enabled by S-stars establishes a comprehensive framework for robustly constraining gravitational theories. As observational capabilities advance with GRAVITY+ and the ELT, S301 is poised to become a cornerstone of gravity tests at large PPN counting parameters, complementary to the Event Horizon Telescope's imaging of the black hole shadow and gravitational wave observations of compact object mergers. The discovery of S301 thus inaugurates a new era in which stellar dynamics and stellar physics combine to probe gravity in a previously unexplored post-Newtonian regime.
	
	\acknowledgments{The Authors sincerely thanks Sunny Vagnozzi for reading the initial draft and for providing valuable technical comments. SC acknowledges the support of Istituto Nazionale di Fisica Nucleare (INFN) Sezioni di Napoli e di Frascati, {\it Iniziative Specifiche} QGSKY and MOONLIGHT2 and the Gruppo Nazionale di Fisica Matematica (GNFM) of Istituto Nazionale di Alta Matematica (INDAM) for the support.}
	
	\appendix
	\section{Derivation of the Combined Uncertainty on $2\gamma-\beta$}\label{A}
	
	The pericenter precession per orbit in the PPN formalism is given by
	\begin{equation}
		\Delta \omega = \frac{6\pi GM}{c^2a(1 - e^2)}\times \frac{2 + 2\gamma - \beta}{3}.
		\label{eq:precession_ppn_deriv}
	\end{equation}
It is convenient to define the dimensionless offset
	\begin{equation}
		X \equiv \frac{2 + 2\gamma - \beta}{3} - 1 = \frac{2\gamma - \beta - 1}{3},
		\label{eq:X_definition}
	\end{equation}
so that $X=0$ in GR and Eq.~(\ref{eq:precession_ppn_deriv}) becomes
	\begin{equation}
		\Delta\omega = \Delta\omega_{\mathrm{GR}}\,(1 + X),
		\label{eq:precession_X}
	\end{equation}
with $\Delta\omega_{\mathrm{GR}} = 6\pi GM/[c^2a(1-e^2)]$. For a single star, the uncertainty on $X$ propagates from the measurement uncertainty $\sigma_{\omega}$ as
	\begin{equation}
		\sigma_{X} = \frac{\sigma_{\omega}}{\Delta\omega_{\mathrm{GR}}}.
		\label{eq:sigma_X_single}
	\end{equation}
	Suppose we have two independent measurements of the same quantity $X$:
	\begin{align}
		X_1 &= X \pm \sigma_{X,1}, \\
		X_2 &= X \pm \sigma_{X,2}.
	\end{align}
	The optimal (minimum-variance) combined estimate is the inverse-variance weighted mean
	\begin{equation}
		X_{\mathrm{comb}} = \frac{X_1 / \sigma_{X,1}^2 + X_2 / \sigma_{X,2}^2}{1 / \sigma_{X,1}^2 + 1 / \sigma_{X,2}^2}.
	\end{equation}
	The uncertainty on this combined estimate propagates as
	\begin{equation}
		\frac{1}{\sigma_{X,\mathrm{comb}}^2} = \frac{1}{\sigma_{X,1}^2} + \frac{1}{\sigma_{X,2}^2}.
	\end{equation}
	Therefore
	\begin{equation}
		\sigma_{X,\mathrm{comb}} = \left( \frac{1}{\sigma_{X,1}^2} + \frac{1}{\sigma_{X,2}^2} \right)^{-1/2}.
	\end{equation}
In our case, $X$ represents the PPN combination $2\gamma-\beta$ (more precisely, the combination $(2\gamma-\beta-1)/3$), and the two measurements come from S2 and S301:
	\begin{equation}
		\sigma_{X,\mathrm{comb}} = \left( \frac{1}{\sigma_{X,\mathrm{S2}}^2} + \frac{1}{\sigma_{X,\mathrm{S301}}^2} \right)^{-1/2}.
		\label{eq:sigma_X_comb}
	\end{equation}
	Substituting Eq.~(\ref{eq:sigma_X_single}) for each star:
	\begin{equation}
		\sigma_{X,\mathrm{comb}} = \left( \frac{\Delta\omega_{\mathrm{S2,GR}}^2}{\sigma_{\omega,\mathrm{S2}}^2} + \frac{\Delta\omega_{\mathrm{S301,GR}}^2}{\sigma_{\omega,\mathrm{S301}}^2} \right)^{-1/2}.
		\label{eq:sigma_X_comb_full}
	\end{equation}
Here $\sigma_{\omega,\mathrm{S2}}$ and $\sigma_{\omega,\mathrm{S301}}$ are the $1\sigma$ uncertainties on the measured precession angles, expressed in radians, and $\Delta\omega_{\mathrm{S2,GR}}$, $\Delta\omega_{\mathrm{S301,GR}}$ are the corresponding GR predictions, also in radians. It is convenient to express the uncertainties as fractions of the GR signal,
	\begin{equation}
		\tilde\sigma_{\omega,i} \equiv \frac{\sigma_{\omega,i}}{\Delta\omega_{i,\mathrm{GR}}},
		\label{eq:tilde_sigma}
	\end{equation}
so that Eq.~(\ref{eq:sigma_X_comb_full}) becomes
	\begin{equation}
		\sigma_{X,\mathrm{comb}} = \left( \frac{1}{\tilde\sigma_{\omega,\mathrm{S2}}^2} + \frac{1}{\tilde\sigma_{\omega,\mathrm{S301}}^2} \right)^{-1/2}.
		\label{eq:sigma_X_comb_norm}
	\end{equation}
For the numerical values quoted in the main text,
	\begin{align}
		\tilde\sigma_{\omega,\mathrm{S2}} &\approx \frac{1.745\times10^{-4}\ \mathrm{rad}}{0.207^\circ \times \pi/180^\circ}
		\approx 4.8\times10^{-2}, \label{eq:tilde_s2}\\
		\tilde\sigma_{\omega,\mathrm{S301}} &\approx \frac{3.491\times10^{-4}\ \mathrm{rad}}{1.95^\circ \times \pi/180^\circ}
		\approx 1.0\times10^{-2}, \label{eq:tilde_s301}
	\end{align}
which, substituted into Eq.~(\ref{eq:sigma_X_comb_norm}), give
	\begin{equation}
		\sigma_{X,\mathrm{comb}} \approx 1.0\times10^{-2}.
		\label{eq:sigma_X_numeric}
	\end{equation}
Finally, since $X = (2\gamma-\beta-1)/3$, we have $2\gamma-\beta = 3X+1$, so
	\begin{equation}
		\sigma_{2\gamma-\beta} = 3\,\sigma_{X,\mathrm{comb}}
		= 3\left(\frac{1}{\tilde\sigma_{\omega,\mathrm{S2}}^2} +\frac{1}{\tilde\sigma_{\omega,\mathrm{S301}}^2}\right)^{-1 / 2}
		\approx 3.0\times10^{-2}.
		\label{eq:sigma_2gamma_minus_beta}
	\end{equation}
This is the combined uncertainty on $2\gamma-\beta$ from the orbital precession measurements of S2 and S301. It agrees with the value $\sigma_{2\gamma-\beta}\approx3.0\times10^{-2}$ quoted in Sec.~\ref{sec:analysis}; the two are now expressed in the same fractional-uncertainty convention of Eq.~(\ref{eq:tilde_sigma}).


\begin{thebibliography}{99}
		
		\bibitem{Will:2014kxa}
		C.~M.~Will,
		Living Rev. Rel. \textbf{17} (2014), 4
		[arXiv:1403.7377 [gr-qc]].
		
		\bibitem{mt1973}
		Misner, C. W., Thorne, K. S., \& Wheeler, J. A. 1973, ``Gravitation'' (San Francisco: W.H. Freeman).
		
		\bibitem{brumberg1991}
		Brumberg, V. A. 1991, ``Essential Relativistic Celestial Mechanics'' (CRC Press).
		
		\bibitem{Hulse:1974eb}
		R.~A.~Hulse and J.~H.~Taylor,
		Astrophys. J. Lett. \textbf{195} (1975), L51-L53
		
		\bibitem{Kramer:2006nb}
		M.~Kramer, I.~H.~Stairs, R.~N.~Manchester, M.~A.~McLaughlin, A.~G.~Lyne, R.~D.~Ferdman, M.~Burgay, D.~R.~Lorimer, A.~Possenti and N.~D'Amico, \textit{et al.}
		Science \textbf{314} (2006), 97-102
		[arXiv:astro-ph/0609417 [astro-ph]].
		
		\bibitem{Damour:1991rd}
		T.~Damour and J.~H.~Taylor,
		Phys. Rev. D \textbf{45} (1992), 1840-1868
		
		\bibitem{Damour:2014tpa}
		T.~Damour,
		Class. Quant. Grav. \textbf{32} (2015) no.12, 124009
		[arXiv:1411.3930 [gr-qc]].
		
		\bibitem{LIGOScientific:2016lio}
		B.~P.~Abbott \textit{et al.} [LIGO Scientific and Virgo],
		Phys. Rev. Lett. \textbf{116} (2016) no.22, 221101
		[erratum: Phys. Rev. Lett. \textbf{121} (2018) no.12, 129902]
		[arXiv:1602.03841 [gr-qc]].
		
		\bibitem{LIGOScientific:2020tif}
		R.~Abbott \textit{et al.} [LIGO Scientific and Virgo],
		Phys. Rev. D \textbf{103} (2021) no.12, 122002
		[arXiv:2010.14529 [gr-qc]].
		
		\bibitem{Bambi:2016sac}
		C.~Bambi, A.~Cardenas-Avendano, T.~Dauser, J.~A.~Garcia and S.~Nampalliwar,
		Astrophys. J. \textbf{842} (2017) no.2, 76
		[arXiv:1607.00596 [gr-qc]].
		
		\bibitem{Krawczynski:2018fnw}
		H.~Krawczynski,
		Gen. Rel. Grav. \textbf{50} (2018) no.8, 100
		[arXiv:1806.10347 [astro-ph.HE]].
		
		\bibitem{Nampalliwar:2019iti}
		S.~Nampalliwar, S.~Xin, S.~Srivastava, A.~B.~Abdikamalov, D.~Ayzenberg, C.~Bambi, T.~Dauser, J.~A.~Garcia and A.~Tripathi,
		Phys. Rev. D \textbf{102} (2020) no.12, 124071
		[arXiv:1903.12119 [gr-qc]].
		
		\bibitem{Abdikamalov:2019hcc}
		A.~B.~Abdikamalov, D.~Ayzenberg, C.~Bambi and S.~Nampalliwar,
		MDPI Proc. \textbf{17} (2019) no.1, 7
		[arXiv:1908.10152 [gr-qc]].
		
		\bibitem{EventHorizonTelescope:2019ggy}
		K.~Akiyama \textit{et al.} [Event Horizon Telescope],
		Astrophys. J. Lett. \textbf{875} (2019) no.1, L6
		[arXiv:1906.11243 [astro-ph.GA]].
		
		\bibitem{EventHorizonTelescope:2022wkp}
		K.~Akiyama \textit{et al.} [Event Horizon Telescope],
		Astrophys. J. Lett. \textbf{930} (2022) no.2, L12
		[arXiv:2311.08680 [astro-ph.HE]].
		
		\bibitem{Volkel:2020xlc}
		S.~H.~V{\"o}lkel, E.~Barausse, N.~Franchini and A.~E.~Broderick,
		Class. Quant. Grav. \textbf{38} (2021) no.21, 21LT01
		[arXiv:2011.06812 [gr-qc]].
		
		\bibitem{Khodadi:2020gns}
		M.~Khodadi and E.~N.~Saridakis,
		Phys. Dark Univ. \textbf{32} (2021), 100835
		[arXiv:2012.05186 [gr-qc]].
		
		\bibitem{Khodadi:2021gbc}
		M.~Khodadi, G.~Lambiase and D.~F.~Mota,
		JCAP \textbf{09} (2021), 028
		[arXiv:2107.00834 [gr-qc]].
		
		\bibitem{Khodadi:2022ulo}
		M.~Khodadi,
		Nucl. Phys. B \textbf{985} (2022), 116014
		[arXiv:2211.00300 [gr-qc]].
		
		\bibitem{Khodadi:2022pqh}
		M.~Khodadi and G.~Lambiase,
		Phys. Rev. D \textbf{106} (2022) no.10, 104050
		[arXiv:2206.08601 [gr-qc]].
		
		\bibitem{Vagnozzi:2022moj}
		S.~Vagnozzi, R.~Roy, Y.~D.~Tsai, L.~Visinelli, M.~Afrin, A.~Allahyari, P.~Bambhaniya, D.~Dey, S.~G.~Ghosh and P.~S.~Joshi, \textit{et al.}
		Class. Quant. Grav. \textbf{40} (2023) no.16, 165007
		[arXiv:2205.07787 [gr-qc]].
		
		\bibitem{Khodadi:2024ubi}
		M.~Khodadi, S.~Vagnozzi and J.~T.~Firouzjaee,
		Sci. Rep. \textbf{14} (2024) no.1, 26932
		[arXiv:2408.03241 [gr-qc]].
		
		\bibitem{Liu:2025wwq}
		W.~Liu, Y.~Liu, D.~Wu and Y.~X.~Liu,
		Phys. Rev. D \textbf{114} (2026) no.2, L021503
		[arXiv:2511.06017 [gr-qc]].
		
		\bibitem{Capozziello:2014rva}
		S.~Capozziello, D.~Borka, P.~Jovanovi{\'c} and V.~B.~Jovanovi{\'c},
		Phys. Rev. D \textbf{90} (2014) no.4, 044052
		[arXiv:1408.1169 [astro-ph.GA]].
		
		\bibitem{Borka:2015vqa}
		D.~Borka, S.~Capozziello, P.~Jovanovi{\'c} and V.~Borka Jovanovi{\'c},
		Astropart. Phys. \textbf{79} (2016), 41-48
		[arXiv:1504.07832 [gr-qc]].
		
		\bibitem{Dialektopoulos:2018iph}
		K.~F.~Dialektopoulos, D.~Borka, S.~Capozziello, V.~Borka Jovanovi{\'c} and P.~Jovanovi{\'c},
		Phys. Rev. D \textbf{99} (2019) no.4, 044053
		[arXiv:1812.09289 [astro-ph.GA]].
		
		\bibitem{Bozza:2010xqn}
		V.~Bozza,
		Gen. Rel. Grav. \textbf{42} (2010), 2269-2300
		[arXiv:0911.2187 [gr-qc]].
		
		\bibitem{Keeton:2005jd}
		C.~R.~Keeton and A.~O.~Petters,
		Phys. Rev. D \textbf{72} (2005), 104006
		[arXiv:gr-qc/0511019 [gr-qc]].
		
		\bibitem{GRAVITY:2018ofz}
		R.~Abuter \textit{et al.} [GRAVITY],
		Astron. Astrophys. \textbf{615} (2018), L15
		[arXiv:1807.09409 [astro-ph.GA]].
		
		\bibitem{Do:2019txf}
		T.~Do, A.~Hees, A.~Ghez, G.~D.~Martinez, D.~S.~Chu, S.~Jia, S.~Sakai, J.~R.~Lu, A.~K.~Gautam and K.~K.~O'Neil, \textit{et al.}
		Science \textbf{365} (2019) no.6454, 664-668
		[arXiv:1907.10731 [astro-ph.GA]].
		
		\bibitem{GRAVITY:2020gka}
		R.~Abuter \textit{et al.} [GRAVITY],
		Astron. Astrophys. \textbf{636} (2020), L5
		[arXiv:2004.07187 [astro-ph.GA]].
		
		\bibitem{GRAVITY:2024tth}
		K.~Abd El Dayem \textit{et al.} [GRAVITY],
		Astron. Astrophys. \textbf{692} (2024), A242
		[arXiv:2409.12261 [astro-ph.GA]].
		
		\bibitem{eldayem2026}
		K.~A.~E.~Dayem, R.~Abuter, N.~Aimar, P.~Amaro-Seoane, A.~Berdeu, J.~P.~Berger, G.~Bourdarot, W.~Brandner, A.~Burkert and D.~Calderon, \textit{et al.}
		[arXiv:2607.12664 [astro-ph.GA]].
		
		\bibitem{GRAVITYPlus:2026}
		R.~Abuter \textit{et al.} [GRAVITY+ Collaboration],
		arXiv:2509.21431 [astro-ph.IM]
		
		\bibitem{Zakharov:2014kka}
		A.~F.~Zakharov, D.~Borka, V.~Borka Jovanovi{\'c} and P.~Jovanovi{\'c},
		Adv. Space Res. \textbf{54} (2014), 1108-1112
		[arXiv:1407.0366 [astro-ph.GA]].
		
		\bibitem{Zakharov:2016lzv}
		A.~F.~Zakharov, P.~Jovanovic, D.~Borka and V.~B.~Jovanovic,
		JCAP \textbf{05} (2016), 045
		[arXiv:1605.00913 [gr-qc]].
		
		\bibitem{DellaMonica:2021xcf}
		R.~Della Monica, I.~de Martino and M.~de Laurentis,
		Mon. Not. Roy. Astron. Soc. \textbf{510} (2022) no.4, 4757-4766
		[arXiv:2105.12687 [gr-qc]].
		
		\bibitem{Rahman:2018fgy}
		M.~Rahman and A.~A.~Sen,
		Phys. Rev. D \textbf{99} (2019) no.2, 024052
		[arXiv:1810.09200 [gr-qc]].
		
		\bibitem{Zakharov:2018omt}
		A.~Zakharov,
		EPJ Web Conf. \textbf{191} (2018), 01010
		[arXiv:1808.05063 [gr-qc]].
		
		\bibitem{Benisty:2023qcv}
		D.~Benisty, J.~Mifsud, J.~Levi Said and D.~Staicova,
		Phys. Dark Univ. \textbf{42} (2023), 101344
		[arXiv:2303.15040 [astro-ph.CO]].
		
		\bibitem{Khodadi:2025upl}
		M.~Khodadi, B.~Pourhassan and E.~N.~Saridakis,
		Phys. Rev. D \textbf{113} (2026) no.6, 064020
		[arXiv:2512.03529 [gr-qc]].
		
		\bibitem{deLaurentis:2022oqa}
		M.~de Laurentis, I.~De Martino and R.~Della Monica,
		Rept. Prog. Phys. \textbf{86} (2023) no.10, 104901
		[arXiv:2211.07008 [astro-ph.GA]].
		
		\bibitem{Losada:2024wdd}
		V.~d.~Losada, R.~Della Monica, I.~de Martino and M.~De Laurentis,
		Astron. Astrophys. \textbf{694} (2025), A280
		[arXiv:2410.22864 [astro-ph.GA]].
		
		\bibitem{DeMartino:2023qkl}
		I.~De Martino, R.~Della Monica and D.~Rubiera-Garcia,
		Phys. Rev. D \textbf{109} (2024) no.2, 024016
		[arXiv:2311.08031 [gr-qc]].
		
		\bibitem{DellaMonica:2023dcw}
		R.~Della Monica and I.~de Martino,
		Phys. Rev. D \textbf{108} (2023) no.10, L101303
		[arXiv:2305.10242 [gr-qc]].
		
		\bibitem{Fernandez:2023kro}
		R.~F.~Fern{\'a}ndez, R.~Della Monica and I.~de Martino,
		JCAP \textbf{08} (2023), 039
		[arXiv:2306.06937 [gr-qc]].
		
		\bibitem{DellaMonica:2022kow}
		R.~Della Monica and I.~de Martino,
		Astron. Astrophys. \textbf{670} (2023), L4
		[arXiv:2206.03980 [gr-qc]].
		
		\bibitem{DellaMonica:2022eeg}
		R.~Della Monica, I.~de Martino, D.~Vernieri and M.~de Laurentis,
		Mon. Not. Roy. Astron. Soc. \textbf{519} (2022) no.2, 1981-1988
		[arXiv:2212.05082 [gr-qc]].
		
		\bibitem{DellaMonica:2021xcf}
		R.~Della Monica, I.~de Martino and M.~de Laurentis,
		Mon. Not. Roy. Astron. Soc. \textbf{510} (2022) no.4, 4757-4766
		[arXiv:2105.12687 [gr-qc]].
		
		\bibitem{Heissel:2021pcw}
		G.~Hei{\ss}el, T.~Paumard, G.~Perrin and F.~Vincent,
		Astron. Astrophys. \textbf{660} (2022), A13
		[arXiv:2112.07778 [astro-ph.GA]].
		
				
		\bibitem{Lechien:2023psa}
		T.~Lechien, G.~Hei{\ss}el, J.~Grover and D.~Izzo,
		Astron. Astrophys. \textbf{686} (2024), A179
		[arXiv:2308.09170 [astro-ph.GA]].
		
		\bibitem{Seoane:2026lyg}
		P.~A.~Seoane, X.~Chen, A.~Torres-Orjuela, R.~Genzel, F.~Eisenhauer, T.~Ott, S.~Gillessen, G.~Bourdarot, D.~C.~Ribeiro and M.~S.~Bordoni, \textit{et al.}
		[arXiv:2607.26134 [astro-ph.GA]].
		
		\bibitem{Iorio:2025ptl}
		L.~Iorio,
		Universe \textbf{11} (2025) no.6, 189
		[arXiv:2506.10162 [gr-qc]].
		
		\bibitem{Capozziello:2011et}
		S.~Capozziello and M.~De Laurentis,
		Phys. Rept. \textbf{509} (2011), 167-321
		[arXiv:1108.6266 [gr-qc]].
		
		\bibitem{Capozziello:2012ie}
		S.~Capozziello and M.~De Laurentis,
		Annalen Phys. \textbf{524} (2012), 545-578
		
		\bibitem{Capozziello:2005bu}
		S.~Capozziello and A.~Troisi,
		Phys. Rev. D \textbf{72} (2005), 044022
		[arXiv:astro-ph/0507545 [astro-ph]].
		
		\bibitem{Sanghai:2016tbi}
		V.~A.~A.~Sanghai and T.~Clifton,
		Class. Quant. Grav. \textbf{34} (2017) no.6, 065003
		[arXiv:1610.08039 [gr-qc]].
		
		\bibitem{McManus:2017itv}
		R.~McManus, L.~Lombriser and J.~Pe{\~n}arrubia,
		JCAP \textbf{12} (2017), 031
		[arXiv:1705.05324 [gr-qc]].
		
		\bibitem{Bolis:2018kcq}
		N.~Bolis, C.~Skordis, D.~B.~Thomas and T.~Z{\l}o{\'s}nik,
		Phys. Rev. D \textbf{99} (2019) no.8, 084009
		[arXiv:1810.02725 [gr-qc]].
		
		\bibitem{Clifton:2018cef}
		T.~Clifton and V.~A.~A.~Sanghai,
		Phys. Rev. Lett. \textbf{122} (2019) no.1, 011301
		[arXiv:1803.01157 [gr-qc]].
		
		\bibitem{Alexander:2007vt}
		S.~Alexander and N.~Yunes,
		Phys. Rev. D \textbf{75} (2007), 124022
		[arXiv:0704.0299 [hep-th]].
		
		\bibitem{Gonzalez-Espinoza:2021nqd}
		M.~Gonzalez-Espinoza, G.~Otalora, L.~Kraiselburd and S.~Landau,
		JCAP \textbf{05} (2022) no.05, 010
		[arXiv:2112.06117 [gr-qc]].
		
		\bibitem{Liodis:2026aka}
		I.~Liodis, G.~Heissel, R.~Mastroioanni, J.~Grover and D.~Izzo,
		Astron. Astrophys. \textbf{710} (2026), A335
		[arXiv:2605.20574 [gr-qc]].
		
		\bibitem{Joyce:2014kja}
		A.~Joyce, B.~Jain, J.~Khoury and M.~Trodden,
		Phys. Rept. \textbf{568} (2015), 1-98
		[arXiv:1407.0059 [astro-ph.CO]].
		
		\bibitem{Bertotti:2003rm}
		B.~Bertotti, L.~Iess and P.~Tortora,
		Nature \textbf{425} (2003), 374-376
		
		\bibitem{Hofmann:2018myc}
		F.~Hofmann and J.~M{\"u}ller,
		Class. Quant. Grav. \textbf{35} (2018) no.3, 035015
		
		\bibitem{Gainutdinov:2020bbv}
		R.~I.~Gainutdinov,
		Astrophysics \textbf{63} (2020) no.4, 470-481
		[arXiv:2002.12598 [astro-ph.GA]].
		
				
		\bibitem{Piran:2026zjs}
		T.~Piran, P.~Amaro-Seoane, B.~Aytac, G.~Bourdarot, A.~Burkert, D.~Calderon, J.~Cuadra, F.~Eisenhauer, R.~Genzel and S.~Gillessen, \textit{et al.}
		[arXiv:2607.24931 [astro-ph.GA]].
		
		\bibitem{Foschi:2026osv}
		A.~Foschi, F.~H.~Vincent, T.~Paumard and G.~Perrin,
		[arXiv:2607.26713 [astro-ph.GA]].
		
		\bibitem{Mirshekari:2013vb}
		S.~Mirshekari and C.~M.~Will,
		Phys. Rev. D \textbf{87} (2013) no.8, 084070
		[arXiv:1301.4680 [gr-qc]].
		
		\bibitem{Pitt:2025tut}
		O.~Pitt and T.~Clifton,
		Class. Quant. Grav. \textbf{43} (2026) no.6, 065009
		[arXiv:2510.08756 [gr-qc]].
		
	\end{thebibliography}
\end{document}